\documentclass[]{bytedance_seed}

\usepackage{amsmath,amssymb}
\usepackage{enumitem}
\usepackage{xspace}
\usepackage{arydshln}
\usepackage{algorithm}
\usepackage{algorithmic}
\definecolor{codegreen}{rgb}{0,0.6,0}

\newcommand{\sysname}{veScale-FSDP\xspace}
\newcommand{\raggedshard}{{\small\texttt{RaggedShard}}\xspace}
\newcommand{\stridedraggedshard}{{\small\texttt{StridedRaggedShard}}\xspace}
\newcommand{\dbuffer}{{\small\texttt{DBuffer}}\xspace}
\newcommand{\deepspeed}{DeepSpeed\xspace}
\newcommand{\megatron}{Megatron\xspace}
\newcommand{\fsdpone}{FSDP1\xspace}
\newcommand{\fsdptwo}{FSDP2\xspace}
\newcommand{\gpt}{GPT-OSS}
\newcommand{\llama}{LLaMA}
\newcommand{\deepseek}{DeepSeek}

\newcommand{\para}[1]{{\bf \noindent #1 \hspace{1pt}}}

\newcommand{\postfigcaption}{\vspace{-0.10in}}

\title{\sysname: Flexible and High-Performance FSDP at Scale}

\author[*]{Zezhou Wang}
\author[*,\dagger]{Youjie Li}
\author[*]{Zhiqi Lin}
\author[*]{Jiacheng Yang}
\author{Cong Xie}
\author{Guanyu Feng}
\author{Zheng Zhong}
\author{Ziyue Huang}
\author{Hongyu Zhu}
\author{Zhi Zhang}
\author[\dagger]{Yanghua Peng}
\author{Xin Liu}

\affiliation{ByteDance Seed}

\contribution[*]{Equal Contribution}
\contribution[\dagger]{Corresponding authors}

\abstract{

\begin{abstract}

Fully Sharded Data Parallel (FSDP), also known as Zero Redundancy Optimizer (ZeRO), is widely used for large-scale model training, because of its memory efficiency and minimal intrusion on model code.  However, existing FSDP systems rely on fixed element-wise or row-wise sharding formats that conflict with block-structured computations. As a result, they struggle to support modern structure-aware training methods, including block-wise quantization and non-element-wise optimizers such as Shampoo and Muon. In addition, today's implementations incur communication and memory overheads that degrade efficiency at the scale of tens of thousands of GPUs. We introduce \sysname, a novel FSDP system that combines \raggedshard, a flexible sharding format, with a structure-aware planning algorithm to deliver both flexibility and performance. \sysname enables zero-copy FSDP communications and natively supports block-wise quantization and non-element-wise optimizers, achieving 5$\sim$66\% higher throughput and 16$\sim$30\% lower memory usage than existing FSDP systems, while scaling efficiently to tens of thousands of GPUs.

\end{abstract}}

\date{\today}
\correspondence{\email{youjie.li@bytedance.com} and \email{pengyanghua.yanghua@bytedance.com}}
\projectpage{\url{https://github.com/volcengine/veScale}}

\begin{document}
\maketitle

\section{Introduction}
\label{sec:intro}

Large language models (LLMs) have become a transformative technology in everyday applications. Driven by the scaling law~\cite{kaplan2020scaling}, LLMs now reach billions of parameters and approach human-level performance in various domains. Training such giant models requires parallelization techniques that distribute the model and optimizer states across thousands of GPUs~\cite{jiang2024megascale}. Among these, Fully Sharded Data Parallel (FSDP)~\cite{zhao2023pytorch,pytorch2024fsdp2,megatron_fsdp}, also known as \deepspeed ZeRO~\cite{rajbhandari2020zero}, is one of the most fundamental techniques. FSDP is often the first choice because of its memory efficiency and flexible data-parallel programming paradigm that is decoupled from model architecture. When additional scaling is needed, FSDP can be combined with other forms of parallelism~\cite{smith2022using,ma2025veomni}.

However, existing FSDP systems struggle to support modern structure-aware training methods. State-of-the-art models use non-element-wise optimizers such as Shampoo~\cite{gupta2018shampoo} and Muon~\cite{jordan2024muon}, and block-wise quantized training as used in DeepSeek-V3~\cite{liu2024deepseek}, all of which require tensor blocks to remain intact under sharding. 
The core limitation is that existing FSDP frameworks shard model states either element-wise~\cite{rajbhandari2020zero,zhao2023pytorch} or row-wise~\cite{pytorch2024fsdp2,megatron_fsdp}, producing sharding boundaries that often misalign with the required block structure. Consequently, either model developers must intrusively modify the model or optimizer code to match tensor boundaries, or system developers must handle complex boundary checks, padding, and additional communication logic.

Beyond inflexibility, existing FSDP systems fall short of  production throughput and memory targets, where we aim to extract every bit of hardware efficiency. GPU Memory is the tighter constraint: in shared clusters, jobs run out of memory or will operate at the memory limit incurring expensive device-side frees, prompting over-provisioning that leaves GPU resources wasted.  These demands become even more critical when scaling training to over 10K GPUs and trillions of parameters. Few existing FSDP systems can scale to this level while maintaining efficiency. \deepspeed ZeRO~\cite{rajbhandari2020zero} pioneered FSDP research but suffers from fragmented AllGather operations~\cite{deepspeed_AG} and inefficient memory management~\cite{fsdp_record_stream}. PyTorch \fsdpone~\cite{zhao2023pytorch} addresses some AllGather inefficiency, but incurs slow ReduceScatter~\cite{fsdp1_reduce} and does not solve memory overhead~\cite{fsdp_record_stream}. PyTorch \fsdptwo~\cite{pytorch2024fsdp2} improves memory management~\cite{per_parameter_shard_rfc} but introduces a high tensor copy overhead. Meanwhile, both \fsdpone and \fsdptwo suffer from slow collectives due to unaligned communication buffers~\cite{wu2025terabyte,nccl16byte}. \megatron-FSDP~\cite{megatron_fsdp} adopts a zero-copy design that avoids \fsdptwo's tensor copy overhead but requires extra padding, increasing both communication and memory costs.

To address these flexibility and performance shortcomings, we present \sysname, a novel FSDP system that preserves the PyTorch-native \texttt{fully\_shard} API while rearchitecting the PyTorch \fsdptwo backend for flexible and high-performance training at scale:

\noindent $\triangleright$ For flexibility, \sysname introduces a novel sharding format, \raggedshard, which supports arbitrary sharding granularity with custom block sizes for structure-aware training, while seamlessly composing with existing PyTorch DTensor sharding formats.

\noindent $\triangleright$ For performance, \sysname introduces a planning algorithm that rearranges \raggedshard tensors to maximize communication efficiency while respecting their desired sharding blocks. We formulate planning as an NP-hard optimization problem and use practical polynomial-time heuristics that achieve high-quality solutions in practice.

\noindent $\triangleright$  \sysname further provides a high-performance primitive, Distributed Buffer (\dbuffer), that backs \raggedshard tensors with slices of a global buffer, enabling zero-copy access, reducing communication overhead, and reducing memory fragmentation via batched memory allocations.

Our extensive evaluations demonstrate that \sysname outperforms all existing FSDP systems on both dense and sparse LLMs across different scales, achieving 5$\sim$66\% higher throughput and 16$\sim$30\% lower memory usage while scaling efficiently to tens of thousands of GPUs. In addition, case studies show that \sysname natively accommodates both non-element-wise optimizers such as Muon~\cite{jordan2024muon} and block-wise quantization methods such as 8-bit Adam~\cite{dettmers8}. \sysname has been deployed in production for most training workloads at ByteDance Seed, and is portable without relying on internal infrastructure.

\raggedshard code is open-sourced at \url{https://github.com/volcengine/veScale}.

\section{Background and Motivation}
\label{sec:background}

\subsection{Structure–Aware Training}
\label{ssec:struct_aware}

Structure-aware training underpins frontier models such as Kimi K2~\cite{team2025kimi}, Gemini~\cite{team2024gemini}, and DeepSeek-V3~\cite{liu2024deepseek}, and is becoming increasingly important. Representative examples include:

\para{Matrix Optimizers.} Matrix-based optimizers such as Shampoo~\cite{gupta2018shampoo} and Muon~\cite{jordan2024muon} can converge faster than AdamW~\cite{wen2025fantastic}. Their updates operate on parameters in their original 2D matrix rather than on independently sharded elements. In practice, this requires gathering each matrix on a selected device before performing the matrix-level update, and then redistributing the update back.

\para{Block-wise Quantization.} Quantizing model weights~\cite{liu2024deepseek} and optimizer states~\cite{dettmers8} is widely used to improve training efficiency. Block-wise quantization is a common technique because per-block scaling factors mitigate the accuracy loss from reduced precision while preserving system efficiency. However, communication-free block-wise quantization requires each quantization block to reside entirely on a single device. If parameters are sharded without respecting block boundaries, devices must exchange scaling-factor metadata, complicating the implementation and reducing the efficiency gains.

\subsection{DTensor and JaggedTensor}
\label{ssec:dtensor}
\label{ssec:jagged}

\begin{figure}[!t]
    \centering
    \includegraphics[width=0.6\linewidth]{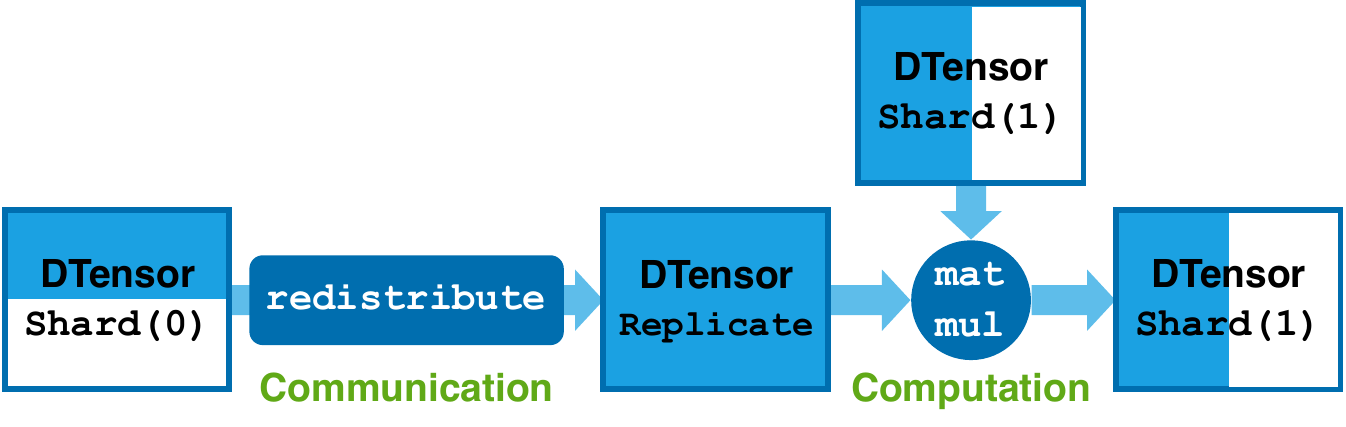}
    \vspace{-1em}
    \caption{Distributed Tensor (DTensor) for flexible communication and computation. The figure shows an example of DTensors executing a sharded matrix multiplication (\texttt{matmul}) on a device. A row-sharded (\texttt{Shard(0)}) DTensor is first redistributed to a replicated (\texttt{Replicate}) DTensor so that it can be multiplied with a column-sharded (\texttt{Shard(1)}) DTensor, producing a column-sharded (\texttt{Shard(1)}) DTensor. The blue part in each DTensor indicates the materialized local tensor on that device.}
    \label{fig:dtensor}
\end{figure}

\para{Distributed Tensor (DTensor)}~\cite{xu2021gspmd, torch-dtensor,li2025vescale} is a promising primitive of PyTorch that provides an opportunity towards structure–aware training. It represents a logical global tensor distributed across multiple devices, with each device holding its local tensor.
DTensor supports three sharding formats (placements): \texttt{Shard(dim)}, which evenly shards a global tensor along a tensor dimension; \texttt{Replicate}, which replicates the global tensor on every device; and \texttt{Partial}, in which each device holds a partial value of the global tensor that must be reduced across devices to materialize the full result. It also provides a \texttt{redistribute} API that converts a DTensor between placements with implicit collective communications. Additionally, DTensors can be computed directly from operators such as \texttt{matmul}, as shown in Figure~\ref{fig:dtensor}. However, a fundamental limitation remains for structure-aware training: the \texttt{Shard} format cannot represent the block-wise sharding needed for quantization or the uneven sharding required by matrix optimizers.

\para{JaggedTensor/NestedTensor on a single device}~\cite{jaggedtensor, nestedtensor, raggedtensors} is a PyTorch/TensorFlow primitive for representing tensors whose last dimension is jagged. For example, a 2D tensor whose rows may have different lengths. These primitives cannot express sharding with block-level granularity, but they offer a useful hint for how \sysname can support structure-aware sharding in distributed training.

\subsection{ZeRO and FSDPs}
\label{ssec:zero_fsdp}

\para{\deepspeed ZeRO}~\cite{rajbhandari2020zero} pioneered this line of FSDP research. Its core idea is to concatenate a layer of tensors (parameters, gradients, and optimizer states) and then shard each concatenated tensor across devices, where tensors may be irregularly partitioned across device boundaries. ZeRO only unshards a layer using AllGather before the forward and backward passes, and reduces the layer gradients using ReduceScatter back across devices. Such a sharding design is fundamentally element-wise and cannot support structure-aware training.

\para{FullyShardedDataParallel (\fsdpone)}~\cite{zhao2023pytorch} is the first PyTorch-native ZeRO, following the same sharding format and limitation, but it is optimized for performance.

\begin{table}[!t]
\caption{Interleaved copy overhead in \fsdptwo for \gpt-120B on 64 H800 GPUs. The table reports the time of interleaved Copy-Out relative to AllGather on the AllGather path, and the time of interleaved Copy-In relative to ReduceScatter on the ReduceScatter path. \texttt{Shard(0)} is the default sharding format for each parameter, while \texttt{Shard(1)} is used when \texttt{Shard(0)} incurs large padding.
}
\label{tab:interleaved_copy_overhead}
\centering
\small
\setlength{\tabcolsep}{4pt}        
\renewcommand{\arraystretch}{0.95} 
\vspace{-0.5em}
\begin{tabular}{lcccc}
\toprule
       & \multicolumn{2}{c}{AllGather path} & \multicolumn{2}{c}{ReduceScatter path} \\
\cmidrule(lr){2-3}\cmidrule(lr){4-5}
       & AllGather & Copy-Out & ReduceScatter & Copy-In \\
\midrule
Shard(0) & 43.71 ms &  5.22 ms & 94.24 ms & 12.37 ms \\
Shard(1) & 44.35 ms & 13.72 ms & 95.36 ms & 23.14 ms \\
\bottomrule
\end{tabular}
\end{table}

\begin{figure}[!t]
    \centering
    \includegraphics[width=0.6\linewidth]{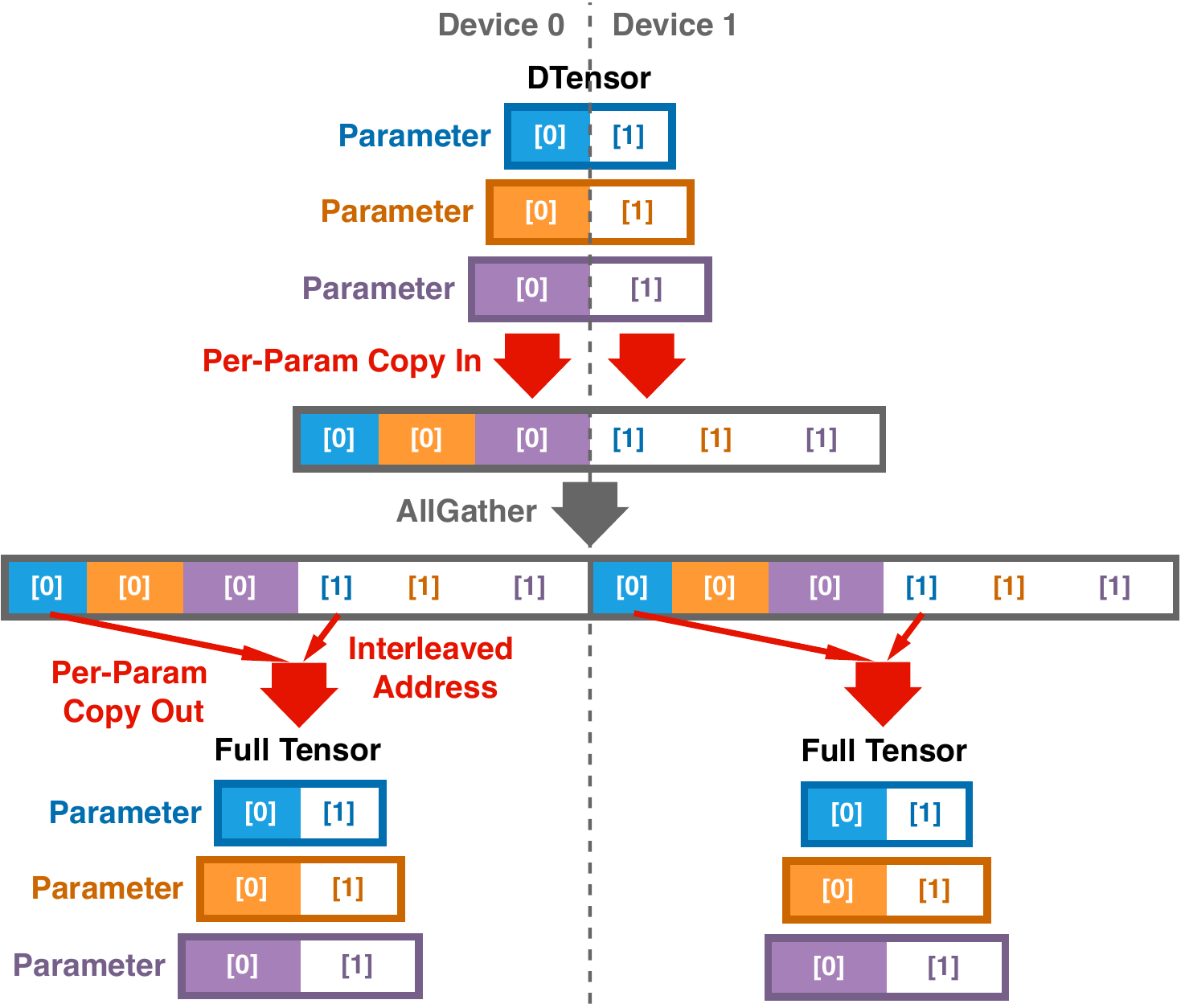}
    \vspace{-0.5em}
    \caption{The copy overhead in the FSDP2 design. Each parameter is represented as a \texttt{Shard(0)} DTensor and is evenly sharded across devices. Before AllGather, the input communication buffer on each device needs to copy in the local shards of all parameters. After AllGather, the output communication buffer stores each parameter in  interleaved memory addresses rather than a contiguous address. As a result, FSDP2 must copy out each parameter from the output buffer to materialize the full tensor with a contiguous address needed for computation. The ReduceScatter path is a reversed one and incurs interleaved Copy-In overhead.}
    \label{fig:fsdp2_copy}
\end{figure}

\para{\texttt{fully\_shard} (\fsdptwo)} is the second PyTorch-native ZeRO and represents the state of the art in the FSDP community. It replaces the concatenated-shard design with per-parameter sharding, representing each tensor as a \texttt{Shard(0)} DTensor. This exposes DTensor flexibility for FSDP parameters in communication, computation, and checkpointing. However, this evenly sharded format still falls short of enabling structure-aware training. Moreover, FSDP2 introduces performance overheads by copying parameters into and from interleaved memory addresses, as shown in Figure~\ref{fig:fsdp2_copy} and Table~\ref{tab:interleaved_copy_overhead}.

\para{\megatron-FSDP}~\cite{megatron_fsdp} is the most recent FSDP prototype optimized for speed. It forgoes \fsdptwo's design and reverts to \fsdpone's concatenated sharding to avoid copying overhead, while further optimizing performance. 
However, \megatron-FSDP introduces a special mechanism to make a concatenation-sharded tensor appear as a \texttt{Shard(0)} DTensor so that model checkpointing can reuse DTensor. 
This mechanism inserts padding into the concatenation so that tensors are sharded row-wise along device boundaries rather than element-wise. Without careful padding planning, the concatenation size can grow significantly, increasing both memory usage and communication volume. Moreover, row-wise sharding still falls short of supporting structure-aware training. Its granularity is fixed by the parameter layout and may not match what the block-wise quantization requires.

\section{Overview}
\label{sec:overview}

\begin{figure}[!t]
    \centering
    \includegraphics[width=0.5\linewidth]{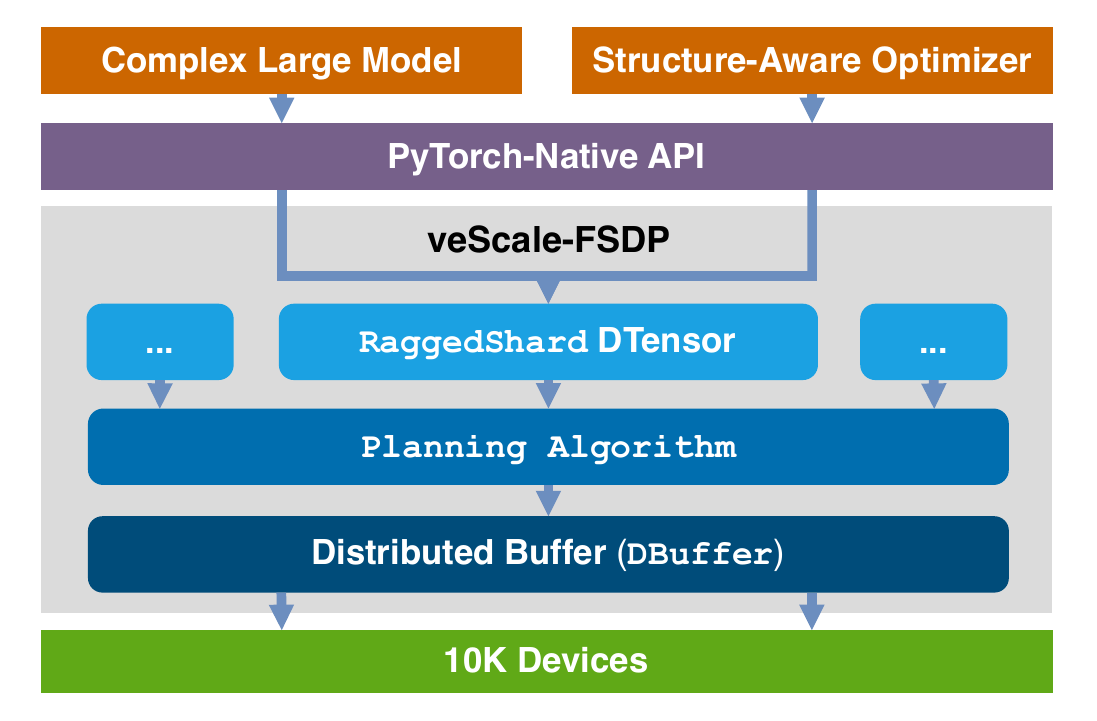}
    \vspace{-1em}
    \caption{\sysname overview.}
    \label{fig:overview}
\end{figure}

To address both flexibility and performance challenges, we present \sysname, a novel FSDP system. Figure~\ref{fig:overview} provides an overview. \sysname allows model developers to build sophisticated large models (e.g., with sparse MoE structures) and structure-aware optimizers (e.g., with non-element-wise operators) that achieve better model quality. At the same time, developers can parallelize these models and optimizers through PyTorch's native \texttt{fully\_shard} API, as in \fsdptwo, without intrusive modifications to model or optimizer code. During parallelization, complex model and optimizer operators can still preserve single-device semantics through a new sharding format,  \raggedshard, which supports arbitrary sharding granularities over contiguous storage and arbitrary distributions across devices for each DTensor (\S\ref{sec:raggedshard}). Under the hood, \raggedshard DTensors are grouped for bucketed communication. To optimize performance, their layouts are rearranged by a planning algorithm derived from an NP-hard optimization problem. The planned layouts are then mapped to a Distributed Buffer (\dbuffer), a new primitive that enables zero-copy communication path with minimal overhead~(\S\ref{sec:dbuffer}). Together, these components enable \sysname to scale efficiently to 10K GPUs in real production deployments.

\section{RaggedShard for Flexibility}
\label{sec:raggedshard}

This section proposes \raggedshard, a novel and general sharding format to increase FSDP’s flexibility for complex models and structure-aware optimizers.

\begin{figure}[!t]
    \centering
    \includegraphics[width=0.6\linewidth]{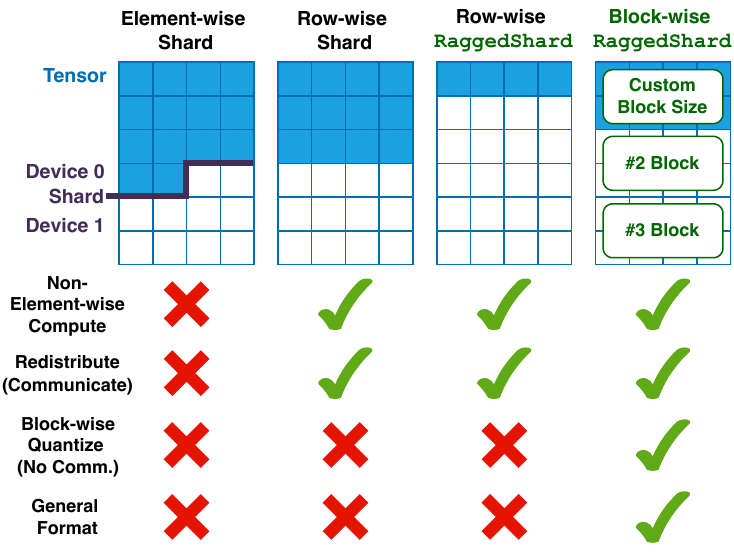}
    \vspace{-1em}
    \caption{Flexibility comparison of different sharding formats.}
    \label{fig:concept_raggedshard}
\end{figure}

\para{Existing sharding formats.}
Figure~\ref{fig:concept_raggedshard} summarizes existing sharding formats for FSDP tensors. The most common and conventional format is the \textit{Element-wise Shard}, where tensors are partitioned arbitrarily across devices without respecting structural boundaries. This leads to dangling elements on each device and a loss of tensor shape and stride information. Consequently, this format severely limits flexibility: (i) non-element-wise operations such as matrix multiplications break on misaligned elements, (ii) collective communication for redistributing tensors across dimensions becomes complex and inefficient, and (iii) quantization methods such as block-wise FP8 quantization cannot align block sizes with random shard boundaries, introducing substantial communication overhead for cross-boundary synchronization. Unfortunately, this element-wise sharding design remains the backbone of \deepspeed and \fsdpone.

The second format is the \textit{Row-wise (Even) Shard}, where a tensor is evenly partitioned along a dimension, with equal-sized shards assigned to each device. This design improves flexibility by enabling non-element-wise computations on sharded tensors and allowing dimension redistribution via \texttt{All2All} collectives. However, it still faces challenges with block-wise quantization, as evenly divided shards are not guaranteed to align with block boundaries. This row-wise sharding format serves as the foundation of \fsdptwo.

\para{The \raggedshard format.}
Inspired by JaggedTensor/NestedTensor~\cite{jaggedtensor, nestedtensor, raggedtensors}, we propose the \raggedshard format for DTensor. \raggedshard increases flexibility by supporting arbitrary \textit{\textbf{sharding granularity}} in contiguous memory, i.e., the size of the atomic non-shardable block, and arbitrary \textit{\textbf{sharding distribution}}, i.e., the number of such blocks placed on each device. The sharding granularity can be defined over contiguous tensor units, such as elements, rows, or higher-dimensional planes. Figure~\ref{fig:concept_raggedshard} shows a simple example: when the sharding granularity is set to one tensor row, \raggedshard yields \textit{Row-wise \raggedshard}, where different devices may hold different numbers of rows. A similar idea has been prototyped in the model checkpointing mechanism of \megatron-FSDP.

The most flexible sharding format is the \textit{Block-wise \raggedshard}, where the sharding granularity is defined as a tensor block with a customizable shape. For example, a tensor may be partitioned into three 2D blocks, with one block placed on device 0 and two blocks on device 1 (see Figure~\ref{fig:concept_raggedshard}). This format not only supports non-element-wise computation and efficient redistribution but also enables block-wise quantization with perfect alignment between quantization blocks and shard boundaries. In fact, the block-wise \raggedshard generalizes all previous sharding formats through different choices of block size.

\begin{figure}[!t]
    \centering
    \includegraphics[width=0.35\linewidth]{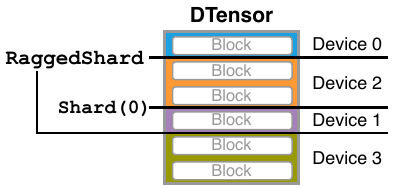}
    \caption{Composability of \raggedshard with existing evenly sharded DTensor placements in 2D parallelism schemes such as FSDP$\times$EP (Expert Parallelism).}
    \label{fig:raggedshard_shard}
\end{figure}

\para{Composing with existing sharding formats.} DTensor has been widely used to express tensor partitions in parallelization strategies such as Tensor Parallelism (TP)~\cite{shoeybi2019megatron} and Expert Parallelism (EP)~\cite{lepikhin2020gshard}. It allows tensors to be represented using replicated, partial-value, or evenly sharded placements along a selected dimension. \raggedshard extends this capability as an additional DTensor placement.

To support combinations of multiple parallelization strategies, \raggedshard needs to compose cleanly with existing DTensor placements. \raggedshard is orthogonal to both replicated and partial-value placements and \sysname specially handles the \texttt{Shard} placement. In practice, TP uses \texttt{Shard(0)} and \texttt{Shard(1)} for column- and row-wise Tensor Parallelism; EP can be encoded as \texttt{Shard(0)} along the expert dimension. By convention, EP/TP is applied before FSDP. In PyTorch, however, the DTensor placement list is organized in the opposite order of conceptual application (see Figure~\ref{fig:raggedshard_shard}): a tensor shown with placements (\raggedshard, \texttt{Shard(0)}) is partitioned as \texttt{Shard(0)} followed by \raggedshard. \sysname reconciles this by: (i) for \texttt{Shard(0)}, introducing a dedicated placement \stridedraggedshard that carries reordering/stride metadata and performs the required reshuffle when materializing the full tensor; and (ii) for \texttt{Shard(dim>0)}, adapting the ragged sharding granularity so it never cuts into that dimension by choosing the granularity as Least Common Multiple (LCM) of the tensor stride of that dimension and user-defined granularity. 

Meanwhile, \raggedshard, as an extended DTensor placement, offers checkpointing capability by directly reusing DTensor-based checkpointing stacks (e.g., PyTorch Distributed Checkpoint~\cite{dcp}) for failure recovery and also inheriting their optimizations such as communication-free sharded checkpointing.

\section{Grouped RaggedShard for Performance}
\label{sec:dbuffer}

This section discusses how to group \raggedshard DTensors for efficient communication. We formulate the underlying optimization problem, prove the NP-hardness of the problem, and present a polynomial-time heuristic algorithm that achieves high-quality solutions in practice.

\begin{figure*}[!t]
    \centering
    \includegraphics[width=\linewidth]{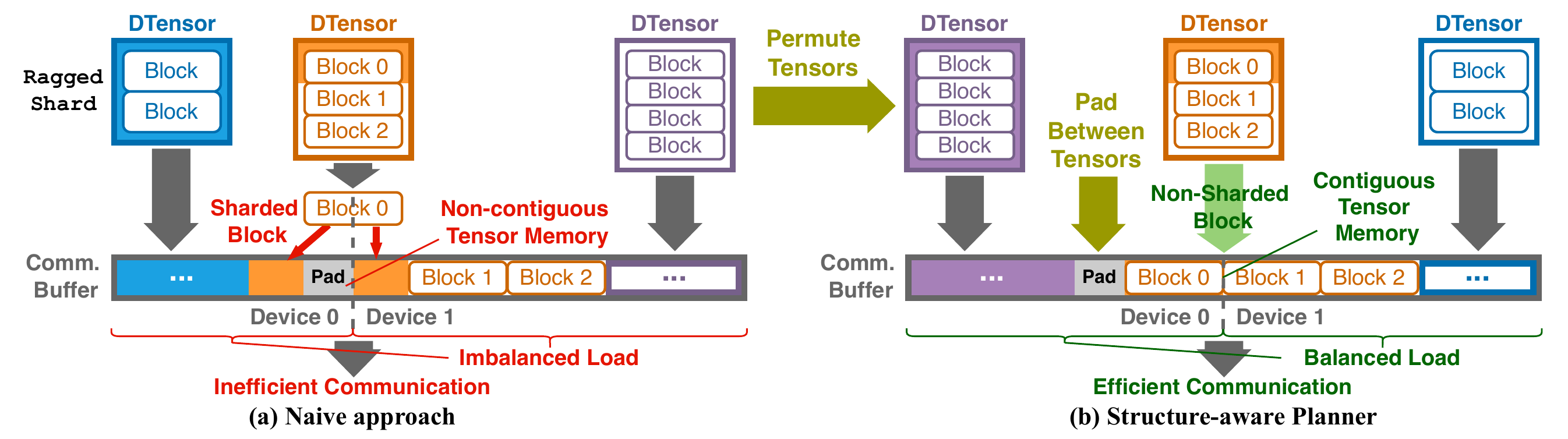}
    \caption{Communication layouts for grouped \raggedshard DTensors. (a) shows a naive strategy that concatenates tensors directly into the communication buffer. This can split blocks across shard boundaries, introduce padding within tensors, and produce unequal per-device buffer sizes, leading to inefficient communication. (b) shows our structure-aware planner that carefully plans the communication layout -- first permutes tensors and then inserts padding between tensors rather than within them. This preserves non-sharded blocks, keeps each tensor contiguous in memory, and balances per-device communication buffers, leading to efficient communication.}
    \label{fig:comm_raggedshard}
\end{figure*}

\para{Challenges for efficient communication.} 
As is well known in the systems community, collective communication relies on tensor bucketing or grouping to maximize network utilization~\cite{li2020pytorch, zhao2023pytorch}. The same principle applies to communication over \raggedshard DTensors in \sysname. However, efficiently grouping \raggedshard tensors is non-trivial and straightforward approaches can lead to significant inefficiencies. Figure~\ref{fig:comm_raggedshard}(a) illustrates three major sources of inefficiency:

\begin{itemize}[leftmargin=*, itemsep=0.25em, topsep=0pt, parsep=0pt, partopsep=0pt]
\item \textit{Sharded block}: Tensors are concatenated into a communication buffer without respecting block boundaries, causing individual blocks to be split across devices; this violates the abstraction of Block-wise \raggedshard and incurs additional communication for quantization.

\item \textit{Non-contiguous tensor memory}: Padding may be inserted at the ends of communication buffers to satisfy collective alignment requirements~\cite{wu2025terabyte,nccl16byte} or equal-size constraints across devices~\cite{nccl2025collective}, but may fall within a tensor. This breaks tensor contiguity and introduces interleaved copy overhead, similar to the Copy-Out after AllGather in Figure~\ref{fig:fsdp2_copy}.

\item \textit{Imbalanced load}: Differences in tensor sizes, block sizes, or padding sizes may lead to unequal buffer sizes across devices. This breaks the symmetry of collective communication and results in underutilized network bandwidth.
\end{itemize}

\para{Towards efficient communication.} 
To efficiently group \raggedshard DTensors, we propose a two-step approach that addresses the above challenges: first permute tensors, and then pad between them rather than padding within individual tensors, as illustrated in Figure~\ref{fig:comm_raggedshard}(b). The key idea is to balance tensor and block sizes across devices while aligning block boundaries in the sharded communication buffer so that blocks are placed contiguously. This approach inevitably introduces some padding overhead, which must be carefully minimized to reduce both memory usage and communication volume.

\para{Optimization problem formulation.} 
Formally, the proposed approach can be formulated as an optimization problem. Let $\mathcal{T} = \{t_1, t_2, \ldots, t_n\}$ denote a set of \raggedshard DTensors, which are sharded across $m$ devices.
Each DTensor $t \in \mathcal{T}$ has a block size of $g_t$, a total tensor size (in elements) $e_t$, and therefore contains $u_t = e_t/g_t$ sharding blocks.
We allocate a global communication buffer and place each $t$ in the buffer as a contiguous memory interval $[\ell_t, r_t)$. The decision variables are a uniform per-device buffer size $S$ and the interval endpoints $\{\ell_t, r_t\}_{t\in\mathcal{T}}$.
The global communication buffer therefore has total size $mS$, partitioned into $m$ equal-sized device-local buffers, where device $k$ owns a memory interval of $[(k - 1)S,\,kS)$ for $k=1,\dots,m$.
Our goal is to minimize $S$ subject to three constraints: Non-Sharded Block, Contiguous Tensor Memory, and Balanced Load (Figure~\ref{fig:comm_raggedshard}(b)):

{
\setlength{\abovedisplayskip}{-3pt}
\begin{equation}
\begin{aligned}
\quad
& \min_{S,\{\ell_t,r_t\}_{t\in\mathcal{T}}} \ S \\[0.25em]
\text{s.t.}\quad
& r_t - \ell_t = e_t\ \land\ r_t \le mS,  \forall\, t\in\mathcal{T}, \\
& r_t \le \ell_{t'} \ \lor\ r_{t'} \le \ell_t, \forall\, t\ne t'\in\mathcal{T}, \\
& kS \le \ell_t \ \lor\ kS \ge r_t \ \lor\ (kS-\ell_t)\equiv 0 \pmod{g_t}, \\
& \hspace{5.5em} \forall\, t\in\mathcal{T},\ \forall\, k=1,\dots,m \\
\end{aligned}\notag
\end{equation}
}

This optimization problem is \textbf{\textit{NP-hard}}, as it can be reduced from the classic Partition problem~\cite{garey1975complexity}.
Although it can be formulated as an Integer Linear Programming (ILP) problem and solved using off-the-shelf solvers, such methods are impractical at scale. In practice, ILP solvers often take tens of minutes to generate a plan and may even trigger system timeouts. Given that user-defined FSDP wrapping can yield hundreds of parameter groups with diverse sharding block sizes, and deployments may span up to hundreds of thousands of devices, we instead design a polynomial-time heuristic algorithm that achieves near-optimal efficiency in practice.

\newcommand{\IndentNo}{\hspace{5em}}
\newcommand{\IndentIn}{\hspace{3em}}
\newcommand{\floor}[1]{\left\lfloor #1 \right\rfloor}
\newcommand{\ceil}[1]{\left\lceil #1 \right\rceil}
\begin{algorithm}[!t]
\caption{Structure-aware planning for grouped communication of \raggedshard DTensors.}
\label{alg:optimalshard}
\begin{algorithmic}[1]
  \STATE \textbf{Input:} ordered tensor list $\mathcal{T}$; per-tensor sharding block size $g_t$ (collectively $G=\{g_t\}$); per-tensor size $e_t$; per-tensor number of blocks $u_t$; number of devices $m$; collective preferred unit size $g_{\mathrm{coll}}$.
  \STATE \textbf{Output:} minimal uniform per-device buffer size $S^\star$.
  \STATE \textbf{Notation:} for a candidate shard size $S$, $dp(t,i;S)$ denotes the minimum number of device-local shards required to place all tensors before $t$ and the first $i$ sharding blocks of tensor $t$.
  \STATE
  \FUNCTION{CheckValidShard($S$)}
      \STATE initialize segment records for $dp(\cdot,\cdot;S)$
      \FORALL{$t \in \mathcal{T}$}
        \STATE $l \gets 0$
        \WHILE{$l < u_t$}
            \STATE // $dp(t,i;S)$ is monotonic in $i$, so skip intermediate states within a constant segment
            \STATE $r \gets \max\{i\!\in\! [l, u_t): dp(t, i; S) = dp(t, l; S)\}$
            \STATE record segment $[l,r]$ with value $dp(t,l;S)$
            \STATE $l \gets r + 1$
        \ENDWHILE
      \ENDFOR
      \STATE \textbf{return} $dp(t_{\mathrm{last}}, u_{t_{\mathrm{last}}}; S) \le m$
  \ENDFUNCTION
  \STATE
  \STATE $g \gets g_{\mathrm{coll}}$
  \STATE $S^\star \gets +\infty$
  \FORALL{$g' \in \textsc{SortAscending}(G)$}
    \STATE $g \gets \textsc{LeastCommonMultiple}(g, g')$
    \STATE $S' \gets \min \{k*g:k\in {N} \land $ CheckValidShard$(k*g)$\}
    \STATE $S^\star \gets \min(S^\star, S')$
  \ENDFOR
  \STATE \textbf{return} $S^\star$
\end{algorithmic}
\end{algorithm}

\para{Heuristic-guided solution.}
\sysname introduces a polynomial-time dynamic-programming (DP) buffer-layout algorithm, guided by permutation heuristics that exploit the regularity of transformer models. The optimization difficulty arises from tensor ordering: in principle, any permutation of tensors could be mapped into the global communication buffer, and finding the global optimum would require exploring all permutations. Fortunately, in practice, transformer parameters are highly structured: linear weights dominate the total parameter count, and sharding blocks are often consistent across layers. To leverage this regularity, we consider three tensor orders: (i) the default tensor order; (ii) sorting by sharding block size; and (iii) sorting by tensor shape. Our statistics show that these orders yield optimal or near-optimal results, so we adopt the default order for simplicity and ease of debugging. For non-transformer architectures, alternative tensor orderings can be plugged in without changing the DP algorithm itself.

Given the tensor order, the proposed algorithm applies a DP procedure to place tensors into the smallest possible global buffer while enforcing the aforementioned three constraints.
It has a time complexity of $O(|\mathcal{T}| ^ 2 m \log(E) \log (|\mathcal{T}| m))$. Algorithm~\ref{alg:optimalshard} presents the details.

The core idea is a case analysis of how each tensor aligns with shard boundaries in any valid layout: (1) it lies entirely within a single local shard; (2) it straddles two adjacent shards but does not fully contain a shard; and (3) it fully contains at least one shard. If every tensor falls into cases (1) or (2), feasibility is monotonic in the shard size $S$: whenever a layout exists for $S$, it also exists for $S+\Delta$, where $\Delta$ is the smallest alignment unit required by the layout. Because every shard includes an inter-tensor boundary, the additional $\Delta$ can always be absorbed as padding. If any tensor falls into case (3), the feasible shard sizes must be multiples of $L=\operatorname{LCM}\{\,g_t \mid \text{$t$ is in case (3)}\,\}$. In this regime, feasibility is monotonic over multiples of $L$: if $kL$ is feasible, then $(k+1)L$ is also feasible. We therefore binary-search for the minimal feasible $S$ over the corresponding multiples, as shown in Lines 21--25. To avoid enumerating an exponential number of candidate case-(3) sets, we sort tensors by element count and consider only prefixes of this sorted order, yielding a 2-approximation. Within \textsc{CheckValidShard}, we define $dp(t,i)$ as the minimum number of devices (shards) required to store all previous tensors and the first $i$ sharding blocks of tensor $t$. Although the DP state space indexes every block position within a tensor, $dp(t,i)$ is monotonic: $dp(t,i)\le dp(t,i+1)$. Therefore, each tensor has at most $m$ distinct DP values. In Lines 10--13, we exploit this property by grouping contiguous indices into segments and skipping intermediate $dp(t,\ast)$ calculations, which yields the stated time complexity.

\begin{figure}[!t]
    \centering
    \includegraphics[width=0.45\linewidth]{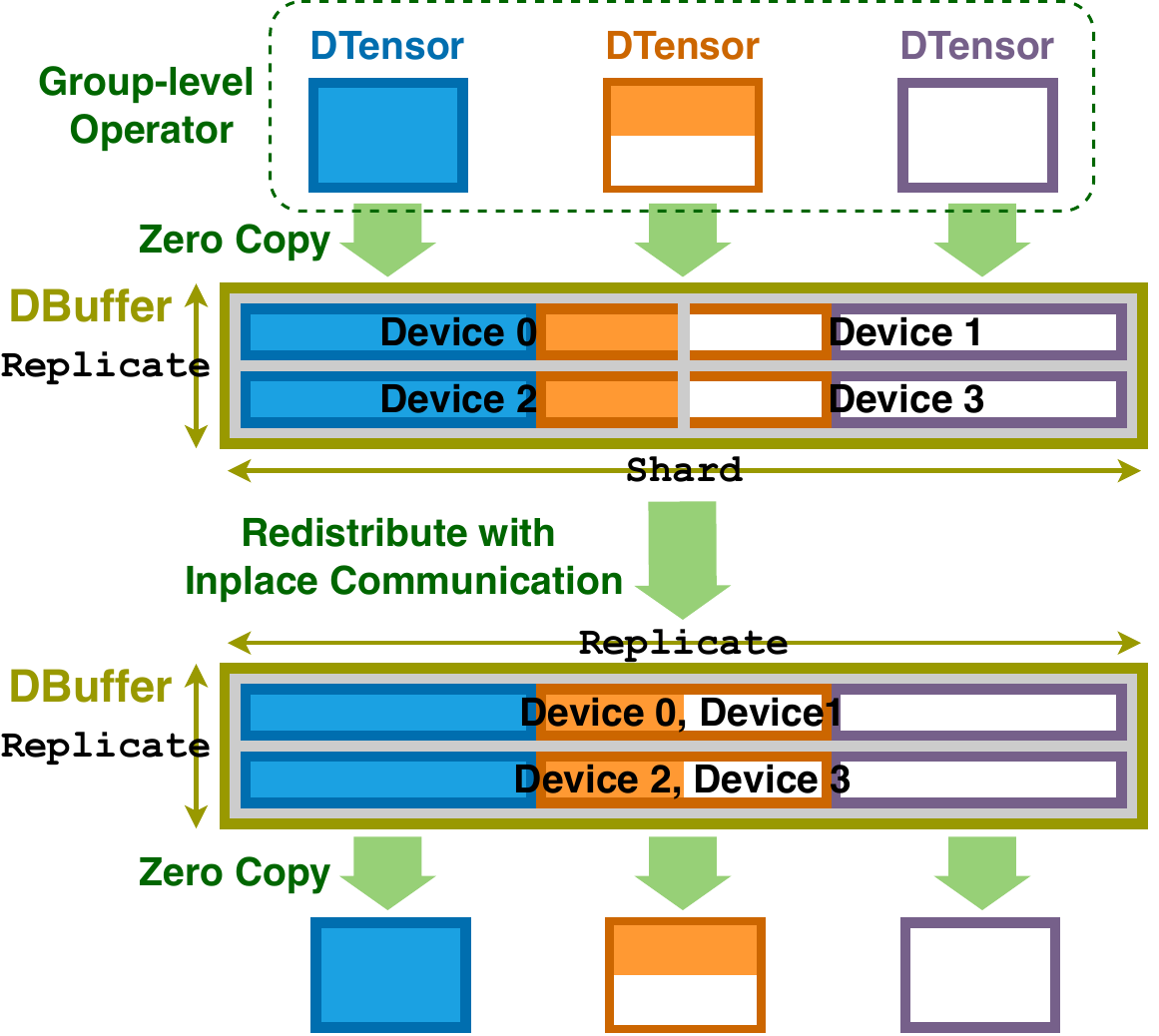}
    \caption{Distributed Buffer (\dbuffer) for high-performance communication. A 2D \dbuffer for parameter AllGather is shown; conversely, a 2D \dbuffer redistributing from \texttt{(Partial, Partial)} to \texttt{(Replicate, Shard)} implements 2D gradient reduction with ReduceScatter and AllReduce.}

    \label{fig:dbuffer}
\end{figure}

\para{Distributed Buffer (\dbuffer)} 
Beyond grouping \raggedshard DTensors, the underlying communication buffer also plays a vital role in achieving high communication, computation, and memory efficiency. \sysname introduces a new primitive, Distributed Buffer (\dbuffer), for efficient grouped DTensors. Figure~\ref{fig:dbuffer} shows the design. 
First, inspired by DTensor, \dbuffer provides global buffer semantics over an $N$-dimensional device topology, with a sharding specification along each dimension, abstracting away the complexity of $N$-D communication and operations.
Second, \dbuffer takes a group of tensors and executes group-level operators rather than per-tensor operators.
For example, before communication, each tensor may need to launch its own CUDA kernels for add, scale, zero, or copy (which may differ across tensors), incurring fragmented compute overhead and blocking communication.
With \dbuffer, identical kernels across tensors are fused before communication, reducing blocking time.
Third, \dbuffer offers zero-copy access before and after communication by leveraging \raggedshard's planning algorithm and providing a persistent address mapping to each tensor's data pointer, thereby minimizing memory footprint and fragmentation. Finally, \dbuffer uses in-place communication and computation.

\vspace{-0.5em}
\section{Evaluation}
\label{sec:eval}

Our evaluation answers the following questions:

\begin{itemize}[leftmargin=*, itemsep=0.25em, topsep=0pt, parsep=0pt, partopsep=0pt]

\item How much does \sysname improve end-to-end training performance over all baseline systems (\S\ref{ssec:end_to_end_perf})?

\item How well does \sysname scale to large device counts (\S\ref{ssec:scalability}), in terms of weak scaling, strong scaling, and model size scaling? 

\item How are 8-bit Adam and Muon optimizer enabled by \sysname's customizable sharding granularity and \raggedshard DTensor, in both performance and development velocity (\S\ref{ssec:muon_optimizer})? 

\item How does the \sysname planner minimize padding, and what is the algorithm overhead (\S\ref{ssec:planning_quality})?

\item How much does each component of \sysname contribute to the training performance (\S\ref{ssec:performance_breakdown})?

\end{itemize}

\textbf{Hardware:} We ran the experiments \S \ref{ssec:end_to_end_perf}, \S\ref{ssec:planning_quality}, and \S \ref{ssec:performance_breakdown} on a NVIDIA H800 cluster; each node contains 8×H800 GPUs (979 BF16 TFLOPS, 80 GB HBM) and 400 GB/s NVLinks. The experiments \S \ref{ssec:scalability}
and \S \ref{ssec:muon_optimizer} were conducted on a NVIDIA Hopper
cluster.

\textbf{Implementation:} \sysname is implemented with 7.6 K lines of code (LoC) in Python, transparently replacing the backend of \fsdptwo while using the same PyTorch-native \texttt{fully\_shard} API.
\sysname serves as a plug-and-play Python module, compatible with standard PyTorch distributed runtimes and a wide range of PyTorch versions.

\textbf{Baselines:} We compare \sysname against state-of-the-art open-source frameworks: \deepspeed ZeRO v0.17.6~\cite{rajbhandari2020zero}, PyTorch~2.7.1 FullyShardedDataParallel (\fsdpone)~\cite{zhao2023pytorch}, PyTorch~2.7.1 \texttt{fully\_shard} (\fsdptwo)~\cite{pytorch2024fsdp2}, and \megatron-FSDP. For fairness, all frameworks are configured to use ZeRO-3 with mixed precision (i.e., FP32 master weights and BF16 forward/backward). Unless otherwise specified, \sysname employs element-wise sharding granularity and is compatible with standard training workflows.

\textbf{Workloads:} For the end-to-end comparison with the baselines (\S\ref{ssec:end_to_end_perf}), we evaluate two state-of-the-art open-source models, \llama-3-70B~\cite{dubey2024llama} and \gpt-120B~\cite{agarwal2025gpt}, as well as an internal MoE model. Under weak scaling, each device is statically assigned one batch; the sequence length is 4096 for the dense \llama\xspace model and 8192 for the MoE models. We use the AdamW optimizer by default. To avoid out-of-memory (OOM) errors for the baselines on \gpt\xspace, we also report results using the SGD optimizer.

\vspace{-0.5em}
\subsection{End-to-End Performance}
\label{ssec:end_to_end_perf}

\setlength{\abovecaptionskip}{0pt}
\begin{figure*}[!t]
    \centering
    \includegraphics[width=\linewidth]{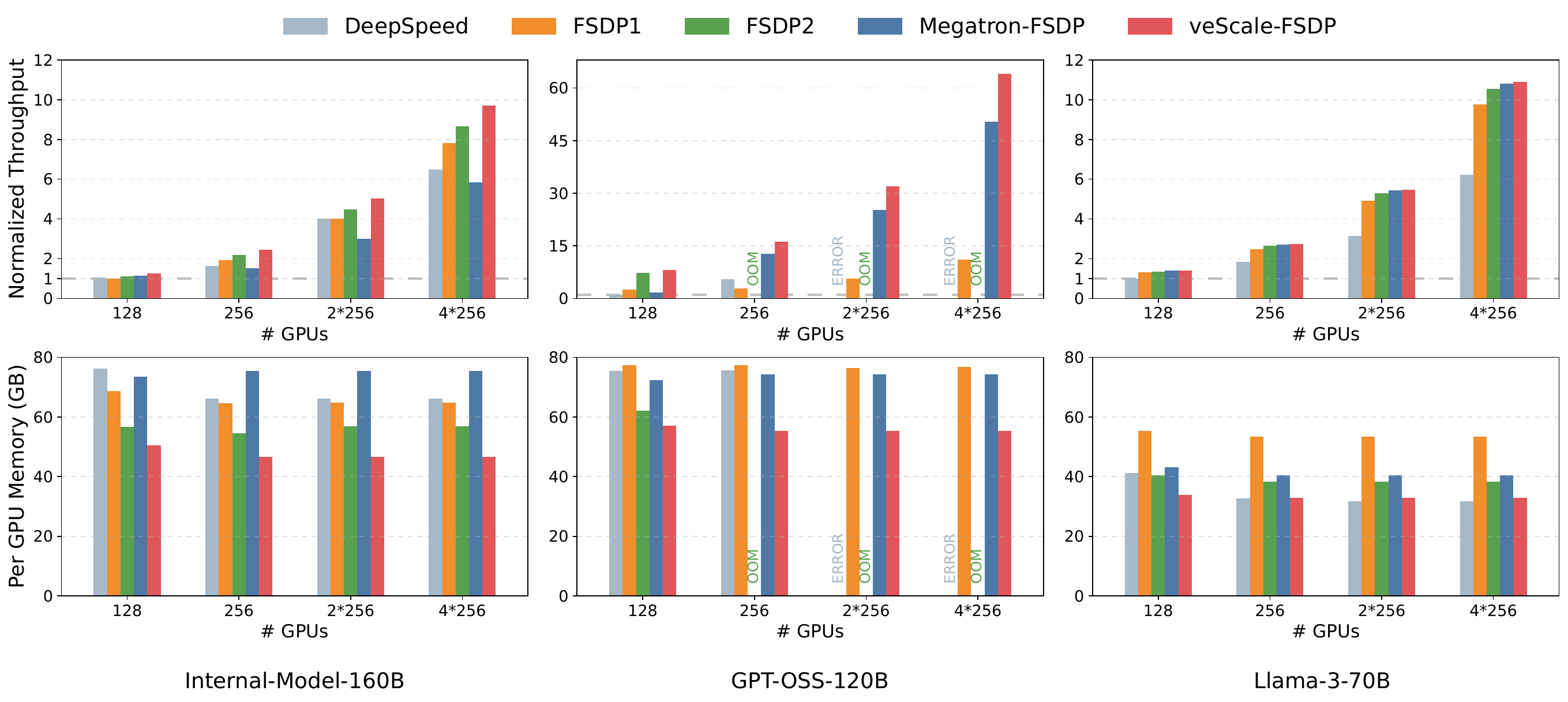}
    \caption{FSDP training performance. 
    Top row: normalized aggregate throughput (tokens/s). Bottom row: peak per-GPU memory (GB). We sweep FSDP (ZeRO-3) at 128/256 GPUs and HSDP with 2- and 4-way replication (2*256, 4*256 GPUs).} 
    \postfigcaption
    \label{fig:zero3_overall_result}
\end{figure*}

Figure~\ref{fig:zero3_overall_result} compares the performance of \sysname against the baselines on the three representative models introduced in the previous section with 1024 GPUs.

\textbf{Throughput:} On the MoE models, \sysname is 11$\sim$66\% faster than all baselines. On \llama-3-70B, \sysname is 5\% faster than \deepspeed, \fsdpone, and \fsdptwo, and slightly ahead of \megatron-FSDP.  The higher throughput arises from optimized communication overlapping, \dbuffer-based zero-copy collectives, and flexible sharding granularities that avoid padding overhead. In contrast, \deepspeed emits fragmented collectives ~\cite{deepspeed_AG}, while \fsdpone exhibits communication bubbles where data movement operations block NCCL progress, under-utilizing the network in both systems. \fsdptwo relies on the per-parameter DTensor even-sharding format that introduces interleaved Copy-Out after AllGather and interleaved Copy-In before ReduceScatter; together these copies can consume up to 14\% of a training iteration and hence reduce throughput. In addition, \fsdpone and \fsdptwo do not enforce NCCL buffer address alignment, so collectives may operate on unaligned memory addresses and suffer substantial communication performance degradation in certain cases~\cite{wu2025terabyte}. Although \megatron is optimized for zero-copy collectives, its fixed \texttt{Shard(0)} sharding granularity, chosen to remain consistent with the upstream DTensor \texttt{Shard(0)} semantics for distributed checkpointing, induces 33\% buffer padding inflation in MoE models and thus slows collective communication~\cite{megatron_fsdp}. Our experiments show that \sysname achieves linear scalability; detailed analysis appears in \S\ref{ssec:scalability}.

\textbf{Memory:} Across benchmarks, \sysname reduces peak \emph{reserved} memory by 16–30\%. The memory saving stems from deterministic, batched \dbuffer memory management: We explicitly manage stream dependencies for predictable memory deallocation, and we batch allocations to reduce fragmentation. By contrast, \deepspeed and \fsdpone inherit non-deterministic deallocations from PyTorch’s implicit \texttt{record\_stream} mechanism~\cite{per_parameter_shard_rfc}, which often prevents the caching allocator from reusing buffers, inflating peak reserved memory by 
20\%. Relative to \fsdptwo’s per-parameter eager allocation, our batched policy yields a further 12\% reduction. \megatron's padding-inflated buffers not only degrade collective efficiency but also raise peak memory by 33\% in MoE experiments; its mixed-precision support persists low-precision buffers, consuming 24\% more memory than \sysname in the \llama-3 experiments. Lower reserved memory translates directly into higher end-to-end efficiency: under high memory pressure, the PyTorch caching allocator issues device frees that synchronize with the driver and stall training. In terms of scalability, \mbox{\sysname's} memory footprint decreases monotonically as the FSDP group size increases and grows only marginally with the replication factor, matching  scaling expectations. A notable exception appears with \gpt: \fsdptwo trains at 128 devices but OOMs at 256. The per-parameter sharding design in DTensor requires padding to enforce even splits along the sharded dimension; with 128 experts spread over 256 devices, the AllGather buffer effectively doubles, exhausting memory. 
While \fsdptwo allows custom sharding along other dimensions, it requires manual padding and thus doubles the interleaved-copy overhead, making it prohibitively expensive (recall Table~\ref{tab:interleaved_copy_overhead}).

\subsection{Scalability and Composability}
\label{ssec:scalability}

\begin{figure*}[!t]
  \centering
  \subfloat[10K-GPU weak scaling\label{fig:weak_scale_10k}]{
    \includegraphics[width=0.24\textwidth]{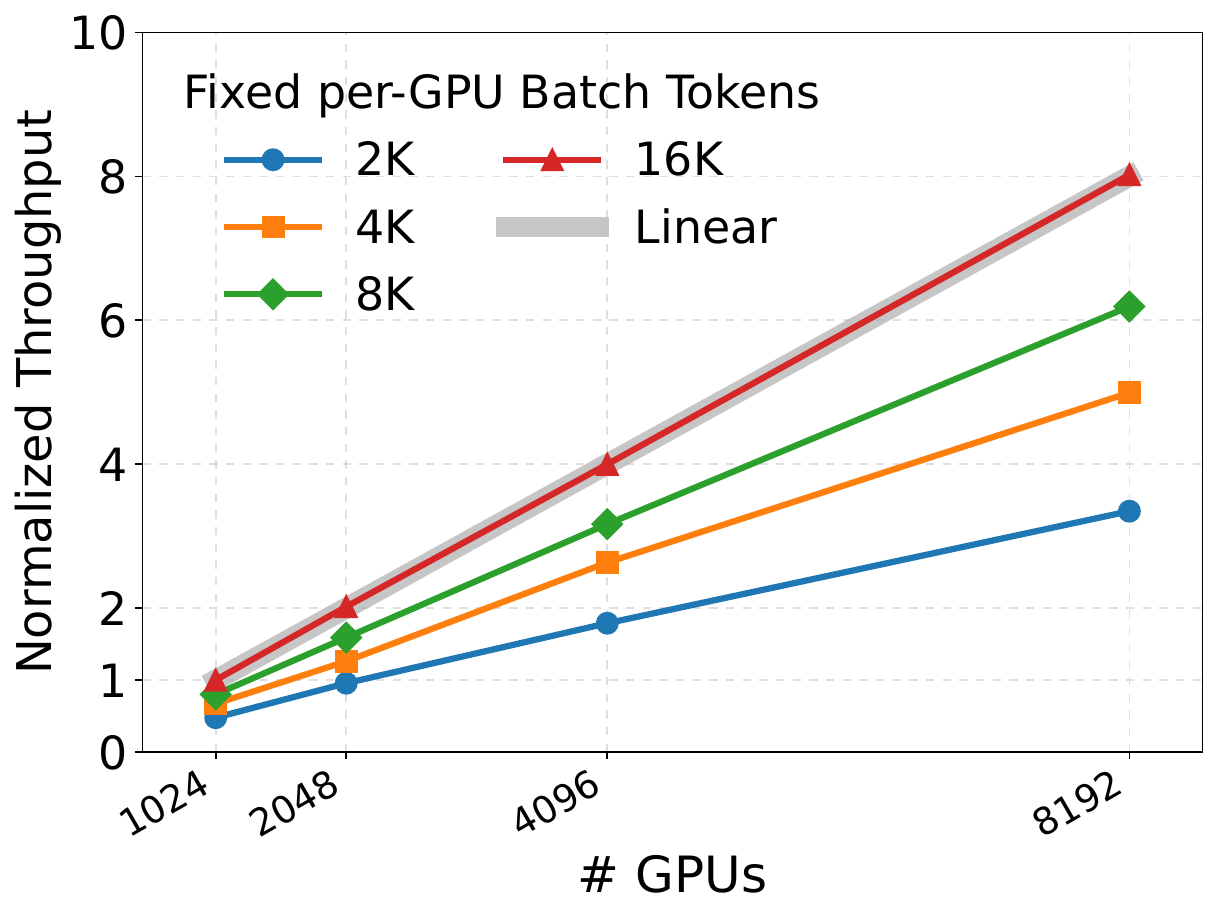}}
  \hfill
  \subfloat[10K-GPU strong scaling\label{fig:strong_scale_10k}]{
    \includegraphics[width=0.24\textwidth]{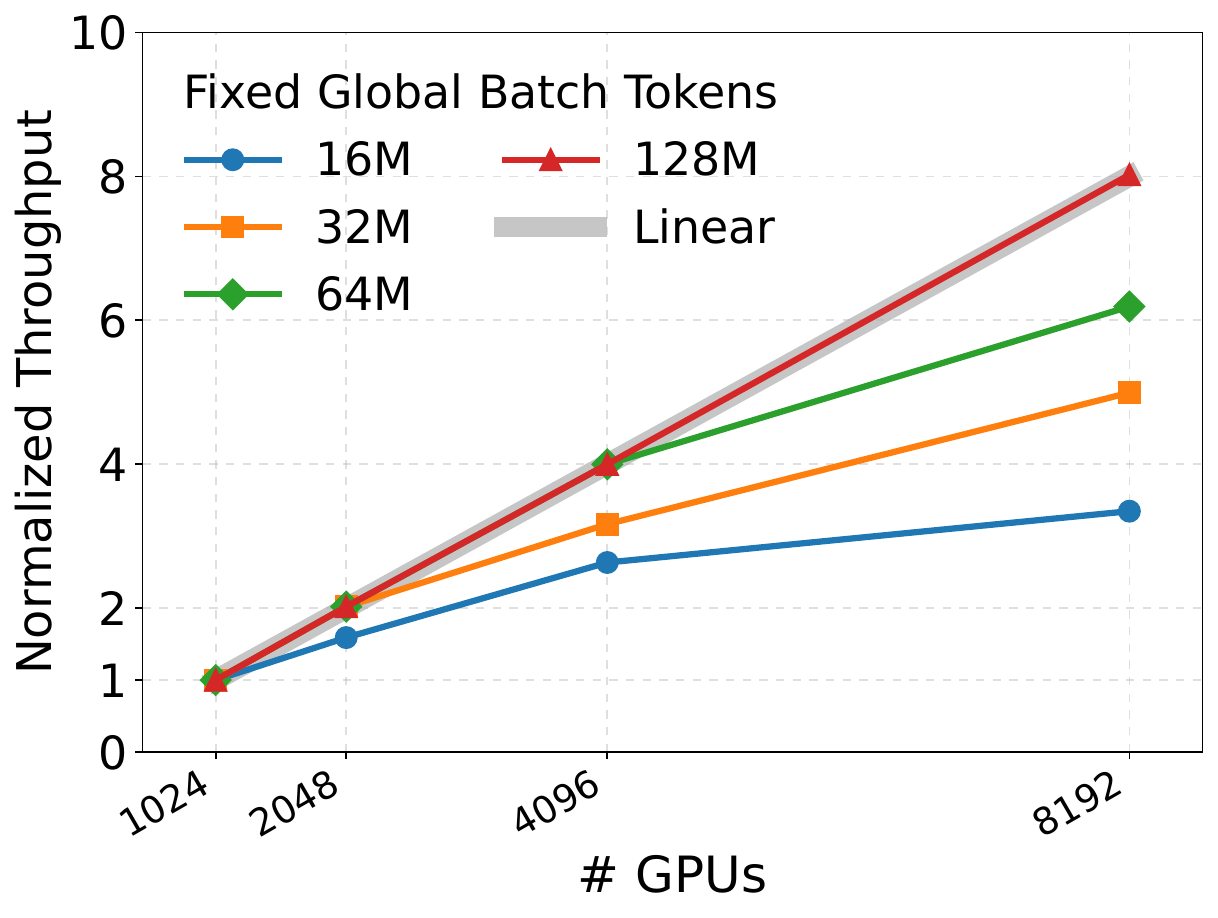}}
  \hfill
  \subfloat[Strong scaling\label{fig:strong_scale_large}]{
    \includegraphics[width=0.24\textwidth]{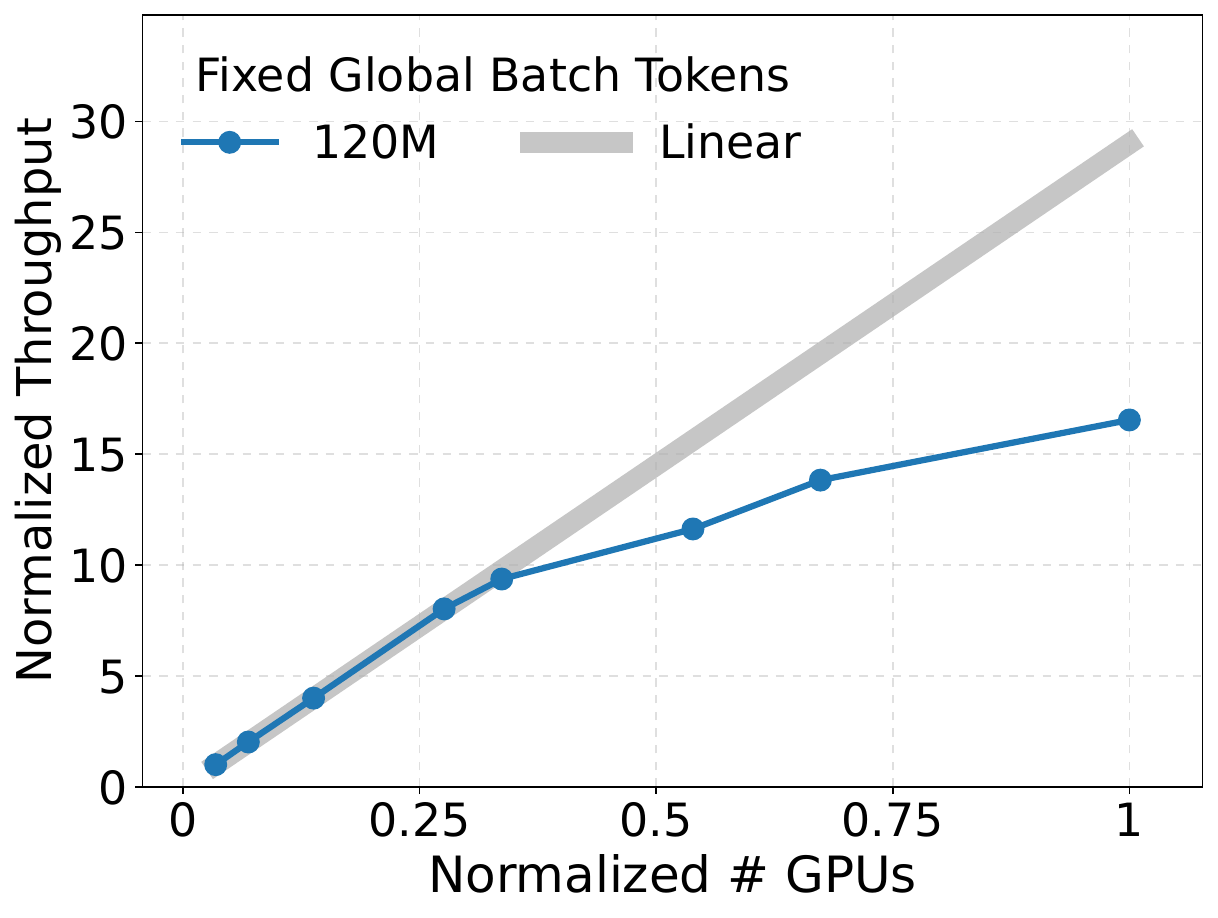}}
  \hfill
  \subfloat[1K-GPU model scaling \label{fig:model_scale}]{
    \includegraphics[width=0.24\textwidth]{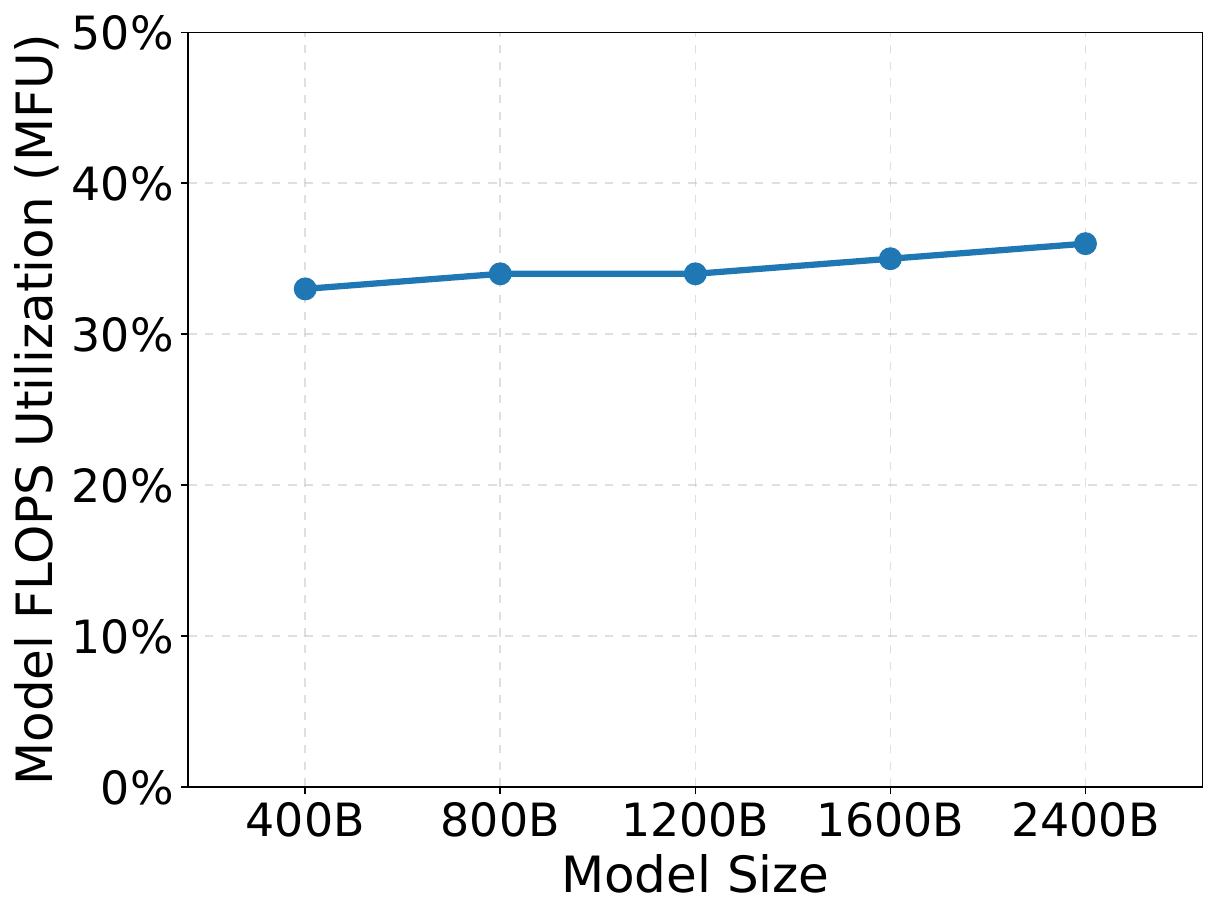}}
  \caption{Scalability of \sysname. (a) Weak scaling to 8K GPUs. (b) Strong scaling to 8K GPUs. (c) Strong scaling at larger scale with normalized throughput and normalized GPU count. (d) Model scaling on 1K GPUs up to 2.4T parameters.}
  \label{fig:scaling_all}
\end{figure*}

The flexibility of \raggedshard also enables seamless integration with complementary parallelization strategies such as Expert Parallelism (EP)~\cite{lepikhin2020gshard}.
Combining these techniques allows \sysname to efficiently scale training to internal models with up to 2.4T parameters on as many as 10K Hopper GPUs, as shown in Figure~\ref{fig:scaling_all}. Note that we evaluate scalability of MoE, because MoE workloads are often more challenging to scale under FSDP: sparse expert computation lowers per-GPU compute while requiring substantial AllGather/ReduceScatter traffic, making communication and padding overheads more significant.

\textbf{Weak scaling:} Figure~\ref{fig:weak_scale_10k} presents the weak scaling performance of \sysname. We train an 800B-parameter MoE internal model on 1K to 8K GPUs while keeping the input size fixed at 2K–16K tokens per GPU. Across all input sizes, \sysname demonstrates near-linear scalability as the GPU count increases. This is expected since the communication cost of FSDP and the computation cost per GPU remain constant with respect to the number of GPUs, depending only on the model and input sizes. These results confirm the efficiency of \sysname on large-scale GPU clusters.

\textbf{Strong scaling:} We further evaluate the strong scaling performance of \sysname by fixing the global batch size to 16M–128M tokens and tuning expert and sequence parallelism configurations for each setting. Figures~\ref{fig:strong_scale_10k} and \ref{fig:strong_scale_large} show the resulting throughput across different numbers of GPUs. \sysname scales linearly with a 120M-token global batch up to 10K GPUs, while still delivering a 3.4$\times$ throughput gain from 1K to 8K GPUs at a 16M-token global batch. Figure~\ref{fig:strong_scale_large} presents the same strong-scaling trend using normalized throughput and normalized GPU count.
When the number of GPUs is small, each GPU processes enough tokens to fully overlap communication with computation, yielding near-linear scaling. However, as the GPU count continues to increase, fewer tokens are assigned per GPU per iteration, causing FSDP communication––including parameter AllGather and gradient ReduceScatter––to dominate runtime. To mitigate this overhead, we adopt cross-node Expert Parallelism, which further reduces FSDP communication time. This optimization introduces higher computation cost due to token exchange and reduced kernel efficiency, resulting in the performance drop at very large scales.

\textbf{Model scaling:} We also evaluate model scaling by fixing the GPU count to 1K and increasing the model size from 400B to 2.4T parameters. With model sparsity constant and 8K training tokens per GPU, we scale both depth (number of layers) and width (intermediate dimensions) proportionally. Figure~\ref{fig:model_scale} reports the effective Model FLOPS Utilization (MFU) per GPU as model size grows. Enabled by efficient memory management of \dbuffer~(\S\ref{sec:dbuffer}), \sysname can train 2.4T-parameter models on only 1K GPUs without any performance degradation. In fact, MFU slightly improves with larger models due to the increased compute intensity and better utilization of GPU resources.

\subsection{8-bit Adam and Muon Optimizer}
\label{ssec:muon_optimizer}

We demonstrate the flexibility of \raggedshard DTensor using two examples: the 8-bit Adam optimizer and the distributed Muon optimizer.

\textbf{8-bit Adam optimizer.} 8-bit Adam~\citep{dettmers8} applies \emph{block-wise} INT8 quantization to the gradient statistics, substantially reducing optimizer-state memory. To enable 8-bit Adam, \sysname exposes an \texttt{orig\_param\_policy} interface that lets users set the quantization \emph{granularity} per parameter. In our setup, we use $32\times 32$ blocks and assign matrix parameters to 32-row block granularity. With this layout, each device quantizes its local shard independently without any communication, and block boundaries are perfectly preserved by \raggedshard. In contrast, existing FSDP systems do not natively track such block boundaries, so enabling block-wise 8-bit Adam often requires intrusive system changes or manual collectives to exchange quantization metadata, incurring both complexity and overhead.

\begin{figure}[!t]
  \centering
  \subfloat[8-bit Adam\label{fig:adam_loss}]{
    \includegraphics[width=0.33\linewidth]{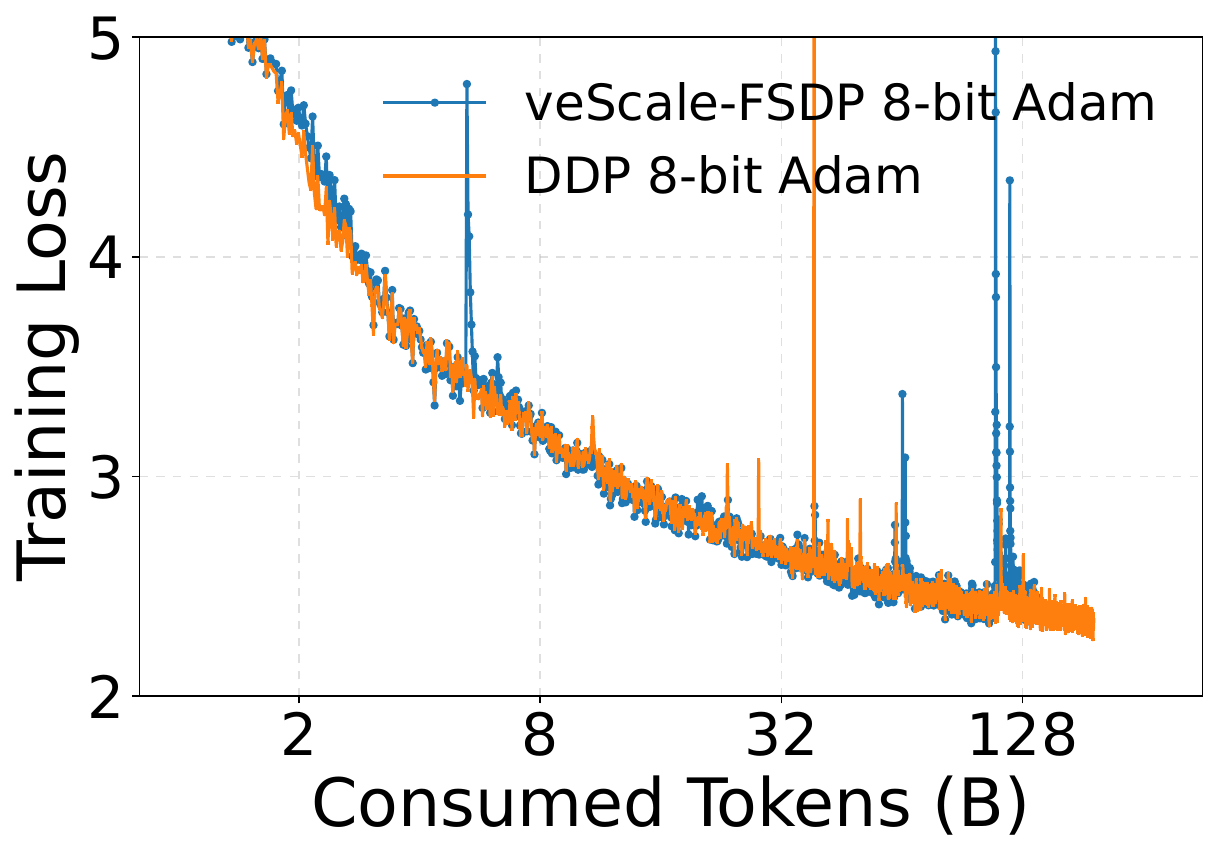}}
  \subfloat[Distributed Muon\label{fig:muon_loss}]{
    \includegraphics[width=0.33\linewidth]{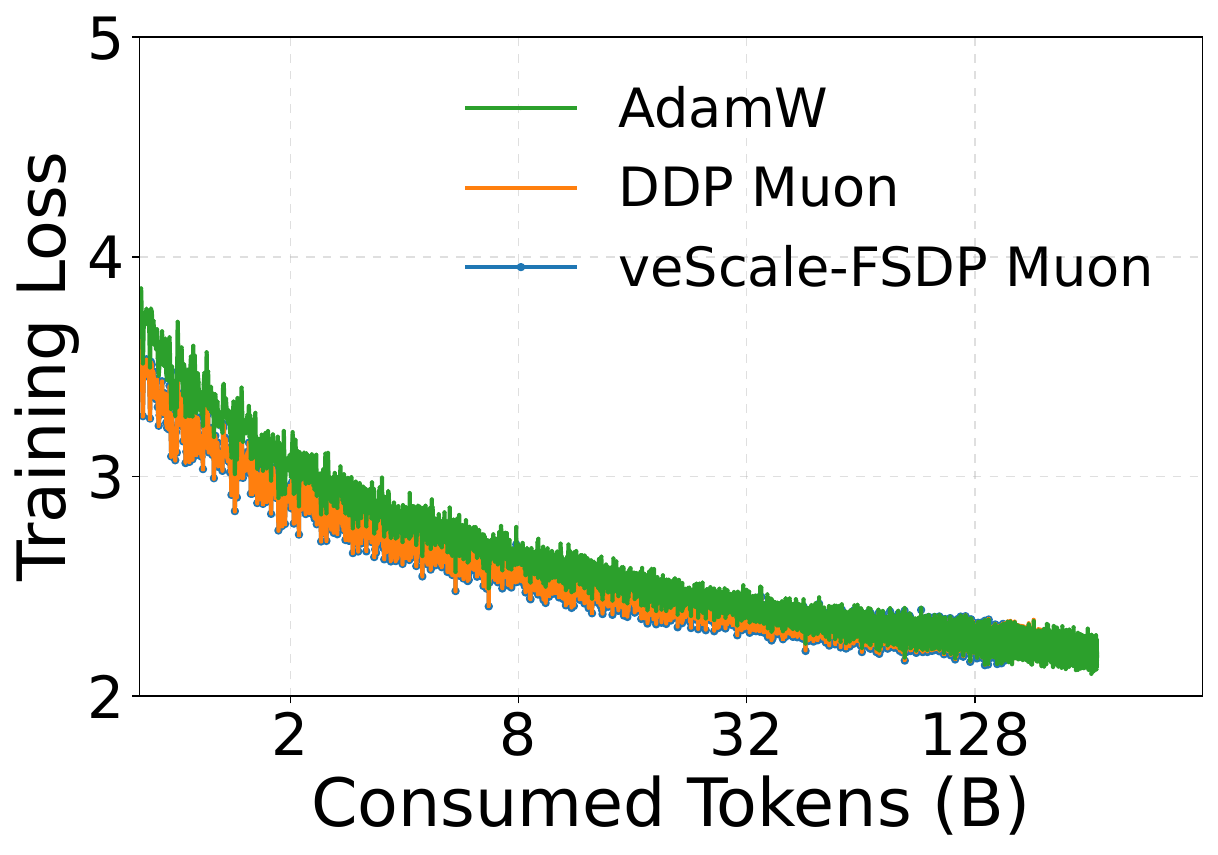}}
  \caption{Training convergence with \sysname on 64 GPUs (FSDP size 64) for 8-bit Adam and distributed Muon.}
  \label{fig:optimizer_loss_both}
\end{figure}

We implement 8-bit Adam using \sysname with few lines of code and provide the evaluation in Figure~\ref{fig:adam_loss}. We compare 8-bit Adam under distributed data parallelism (DDP) and with \sysname. The loss curves track closely, with occasional spikes characteristic of reduced-precision optimizer states. The small difference stems from the gradient-reduction schedule: DDP uses bucketed AllReduce, whereas \sysname performs layer-wise ReduceScatter. (Note that the loss curves in Figure~\ref{fig:optimizer_loss_both} are not directly comparable: for 8-bit Adam, we use a smaller learning rate to mitigate overflow/underflow in reduced precision.)

\begin{algorithm}[!t]
\caption{\raggedshard Distributed Muon}
\label{alg:dist_muon}
\begin{algorithmic}[1]
\FORALL{$\mathbf{w}$ in 2D parameter tensors}
  \STATE $\mathbf{g} \gets \mathrm{grad}(\mathbf{w})$
  \STATE $\mathbf{u} \gets \textsc{MomentumUpdate}(\mathbf{g}, \mathbf{m})$
  \STATE $p \gets \mathrm{placement}(\mathbf{u})$ // \textit{original DTensor placement}
  \STATE // \textit{Choose compute device via load balancing}
  \STATE $r \gets \textsc{SelectRoot}()$
  \STATE // \textit{Unshard to root via redistribute}
  \STATE $\mathbf{o} \gets \textsc{Redistribute}(\mathbf{u}, \raggedshard(r))$
  \STATE // \textit{Muon update: Newton--Schulz on full tensor.}
  \STATE $\mathbf{o} \gets \textsc{NewtonSchulz}(\mathbf{o})$
  \STATE // \textit{Redistribute update back.}
  \STATE $\mathbf{o} \gets \textsc{Redistribute}(\mathbf{o}, p)$
  \STATE $\mathbf{w} \gets \mathbf{w} - \eta\,\mathbf{o}$
\ENDFOR
\end{algorithmic}
\end{algorithm}

\textbf{Distributed Muon optimizer.} The matrix-sign preconditioner (e.g., Newton–Schulz) of Muon requires the \emph{full} 2D parameter matrix with its original shape. Algorithm~\ref{alg:dist_muon} sketches the distributed Muon optimizer enabled by \raggedshard. Thanks to \raggedshard's capability to support uneven sharding, users can write Muon’s parameter-gather step in a clean SPMD way: after redistribution, only the root rank holds the full 2D parameter, so the Newton–Schulz update becomes a no-op on other ranks. As lines 5–8 show, the algorithm selects a root via load balancing and unshards to it using the standard DTensor \texttt{redistribute} with \raggedshard placement. Lines 9–10 run the Muon matrix iteration only on the root that holds the full tensor. Finally, lines 11–13 redistribute the update back to the original device and apply it. Therefore, users do not need to handle the complex logic of communication and can further overlap communication with computation via asynchronous \texttt{redistribute}. In addition, our optimized Muon reaches 47.3\% MFU on 256 Hopper GPUs by exploiting the communication-computation overlapping and using \texttt{torch.compile} to further increase compute density.

We implement distributed Muon using \sysname with few lines of code and provide the evaluation in Figure~\ref{fig:muon_loss}.
We compare the loss curves of Muon with AdamW: the two Muon runs (\sysname and DDP) match closely, and Muon converges faster than AdamW, stabilizing around 0.01 lower loss after training $\sim$80B tokens, which is consistent with prior results~\citep{wen2025fantastic}.

\vspace{-0.5em}
\subsection{Planning Quality}
\label{ssec:planning_quality}

A major design objective of the planning algorithm (Algorithm~\ref{alg:optimalshard}) is to enable arbitrary granularity of \raggedshard, while minimizing padding overhead and thus reducing communication volume.
The quality of the planning algorithm can be directly evaluated by padding size.
We evaluate it by benchmarking \deepseek-V3-671B and \gpt-120B across varying device counts. Following a \deepseek-style quantization scheme, we quantize only the FFN weights (most parameters) and sweep the row granularity of expert-MLP matrices over {128, 16, 1}. The 128-row setting reproduces \deepseek's 128$\times$128 tiling (i.e., weights can be sliced into 128$\times$128 blocks). We then report the resulting relative padding ratios and analyze the root cause of the extra padding.

\begin{figure}[!t]
  \centering
  \subfloat[\deepseek-V3-671B\label{fig:deepseek_v3_padding}]{
    \includegraphics[width=0.30\linewidth]{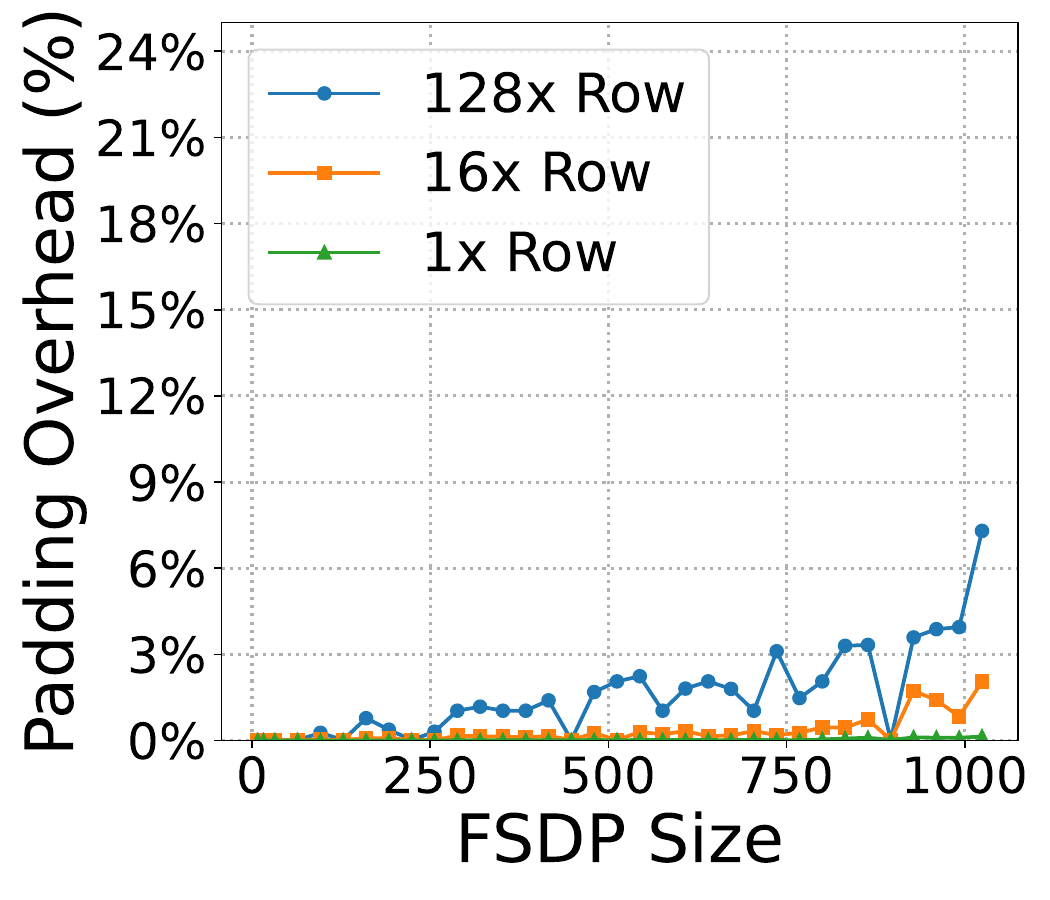}}
  \subfloat[\gpt-120B\label{fig:gpt_oss_padding}]{
    \includegraphics[width=0.30\linewidth]{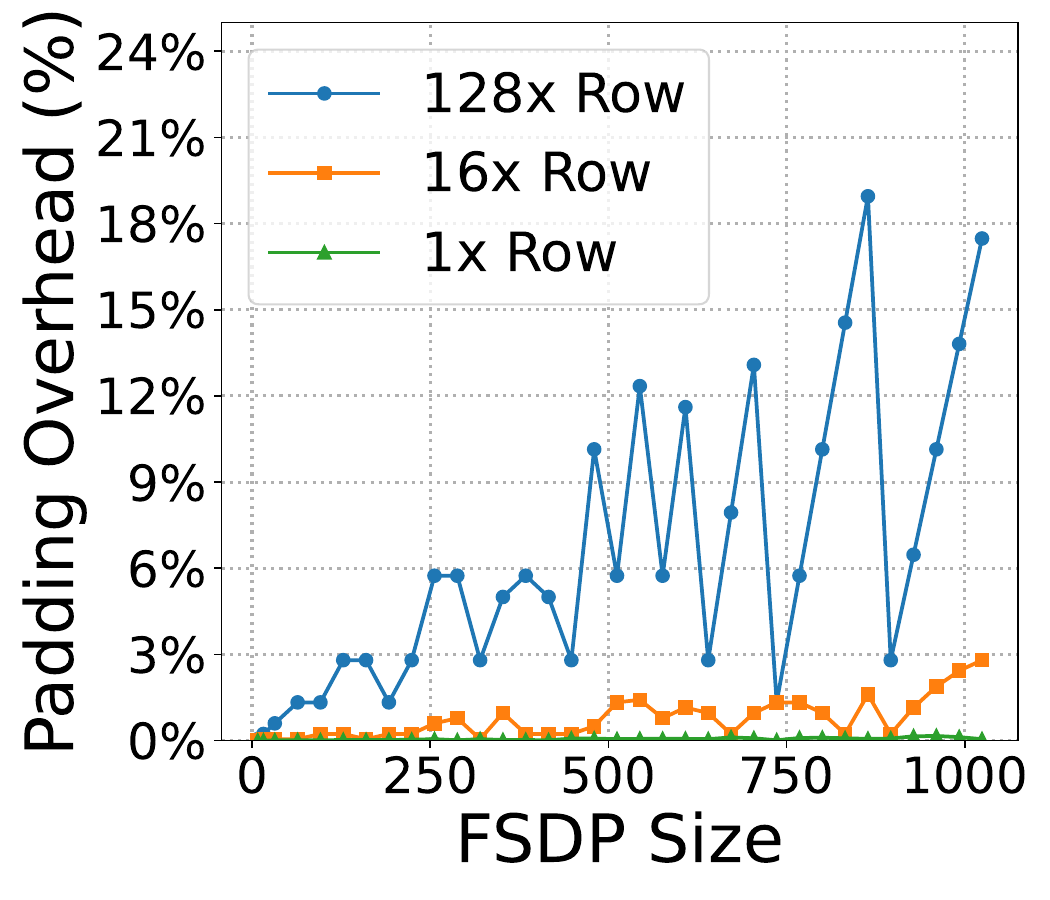}}
  \caption{Padding overhead (extra padding bytes over total parameter size) of \raggedshard communication.
  Lines compare the sharding granularities (1$\times$/16$\times$/128$\times$ parameter row size) versus FSDP sharding size (number of GPUs).}
  \label{fig:padding_ratio_both}
\end{figure}

Figure \ref{fig:deepseek_v3_padding} and \ref{fig:gpt_oss_padding} show that with 1$\times$ and 16$\times$ row granularities, \sysname keeps padding overhead less than 3\% across all FSDP sizes for both models. With 128$\times$ rows, \deepseek-V3 remains mostly below 3\% with mild growth, whereas \gpt\xspace exhibits step-like fluctuations with spikes up to 18\%. \gpt\xspace fuses all experts into a single parameter tensor, whereas \deepseek-V3 materializes each expert as a separate parameter; this enables per-expert padding between MLPs and thus relaxes the global padding constraint. The fluctuation behavior is expected: each matrix must be partitioned across the shard group in discrete quanta determined by (i) the granularity unit (e.g., rows) and (ii) NCCL’s even-input alignment for high-performance collectives. Effective shard sizes are therefore rounded up to the least common multiple of these granularities; when the group size crosses a multiple, the per-device shard size jumps, producing the observed spikes.

We also evaluate the overhead of the planning algorithm itself: the algorithm runtime is less than 0.3 seconds across all experiments, which is one-time and negligible in distributed training initialization.

Based on Figure~\ref{fig:padding_ratio_both} and our experience, a practical guideline is to avoid excessively large FSDP group sizes and use hierarchical parallelism (e.g., HSDP) to scale to larger GPU counts. In practice, we select the FSDP group size by offline simulation to minimize LCM-induced rounding for a given model and \raggedshard granularity. When model dimensions remain configurable, choosing hidden sizes divisible by small composite factors rather than large co-prime dimensions can further reduce padding overhead.

\subsection{Performance Breakdown}
\label{ssec:performance_breakdown}

\begin{table}[!t]
\caption{Component ablation for 8-bit Adam, reported as normalized throughput after disabling each component independently. N/A indicates that the corresponding configuration is not meaningfully runnable without intrusive modifications to the model or optimizer code or manual management of custom collectives.}
\label{tab:8_bit_adam_ablation}
\centering
\small
\setlength{\tabcolsep}{4pt}        
\renewcommand{\arraystretch}{0.95} 
\vspace{0.5em}
\begin{tabular}{ll}
\toprule
\textbf{\sysname Component} & \textbf{Normalized Throughput} \\
\midrule
Combined & 100.0\% \\
Disable \dbuffer only & 92.8\% \\
Disable Planning Algorithm only & 65.4\% \\
Disable \raggedshard only & N/A \\
\bottomrule
\end{tabular}
\end{table}

To quantify the benefit of each component, we ablate \sysname by disabling one component at a time and report the resulting throughput, normalized to the full system. We run this study on 32 GPUs when training a \gpt-style model with 8-bit Adam.

Table~\ref{tab:8_bit_adam_ablation} shows that \dbuffer and the planning algorithm account for most of the realized speedups: disabling them reduces throughput to 92.8\% and 65.4\%, respectively. In contrast, \raggedshard is not just an optimization; it is the abstraction that makes block-wise 8-bit Adam usable without intrusive model/optimizer changes or hand-written collectives. Specifically:

\begin{itemize}[leftmargin=*, itemsep=0.25em, topsep=0pt, parsep=0pt, partopsep=0pt]

\item \textbf{\dbuffer.} Disabling \dbuffer drops throughput by 7.2\%, reflecting the Copy-In/Copy-Out overhead around collectives when communication buffers require copying.

\item \textbf{Planning Algorithm.} Disabling the planning drops throughput by 34.6\% because quantization blocks are no longer guaranteed to be fully contained within a device's local shard. The system then falls back to DTensor redistribution to assemble the required optimizer states before per-block quantization, incurring substantial extra communication overhead.

\item \textbf{\raggedshard DTensor.} Disabling \raggedshard makes it effectively non-runnable: users must either (i) carefully change every model and optimizer tensor so that $32\times 32$ block boundaries align with shard boundaries, or (ii) manually implement complex collectives (e.g., per-block metadata exchange and state gathering) to recover block-wise semantics. We therefore report this setting as N/A to indicate it is not meaningfully usable.

\end{itemize}

\vspace{-0.5em}
\section{Lessons Learned}
\label{sec:lesson}

During the deployment of \sysname for real industrial workloads that use more than 10K GPUs, we summarize the key lessons we have learned.

\vspace{-0.15em}
\textbf{Lesson-1: Small-scale workloads can predict large-scale performance.} The performance of FSDP-based workloads can be accurately estimated using each layer’s computation time and FSDP communication time. Computation occurs entirely within each GPU, and FSDP communication time remains largely unchanged when the number of GPUs increases. This observation is validated by our weak scaling experiments~(\S\ref{sec:eval}). In practice, we profile the performance of \sysname on around 64 GPUs and extrapolate to thousands of GPUs, achieving similar results.

This extrapolation assumes that the profiling run exercises network behavior similar to the target scale: comparable network topology, identical collective algorithms/protocols, and a sufficiently large workload to reach bandwidth saturation. To further improve predictability at large scales, we use additional parallelization (e.g., HSDP/EP) to cap the collective group size, preventing excessively large collectives whose latency can vary more.

\vspace{-0.15em}
\textbf{Lesson-2: Design system abstractions on the shoulders of giants.} DTensor provides a powerful abstraction that already supports a wide range of parallelization techniques. By designing new abstractions on top of DTensor, we can seamlessly integrate existing parallelization strategies. In our work, \raggedshard is implemented as an optional placement on DTensor, enabling easy collaboration of established infrastructure such as Tensor and Expert Parallelism, as well as mature training tools like distributed checkpointing~\cite{dcp}. This approach minimizes engineering effort while contributing to a shared ecosystem that benefits the broader community. In fact, \raggedshard has already appeared as a planned feature on the official roadmap~\cite{pytorch_2026_h1_roadmap} of PyTorch.

\vspace{-0.15em}
\textbf{Lesson-3: Decoupled model definition with system optimization matters.} The rapid evolution of model architectures demands frequent updates to model definitions. However, existing frameworks such as Megatron-LM tightly couple system-level parallelization optimizations with model code, making it difficult for researchers to modify or extend architectures. In \sysname, we decouple model definition from the system framework, allowing researchers to focus on model design while maintaining linear scalability across up to 10K GPUs. This separation greatly simplifies model development and accelerates architectural innovation.

\vspace{-0.5em}
\section{Conclusion}
\label{sec:conclusion}

\sysname is a scalable training system that combines high flexibility with high performance through the \raggedshard abstraction and a structure-aware planning algorithm that maximizes GPU utilization. Experiments demonstrate that \sysname seamlessly integrates with emerging techniques such as Muon optimizers and significantly outperforms existing systems, achieving 5$\sim$66\% higher throughput and 16$\sim$30\% lower memory usage, while scaling efficiently to tens of thousands of GPUs.

\section{Acknowledgments}
\label{sec:acknowledge}
\sysname would not have been possible without the tremendous support and collaboration of our teammates and colleagues. We sincerely thank them (in no particular order; this list is not exhaustive):

\begin{itemize}[leftmargin=*, itemsep=0.25em, topsep=0pt, parsep=0pt, partopsep=0pt]

\item veScale members: Hongrui Zhan, Ziyi Zhang, Hao Feng
\item ByteDance teammates: Jianyu Jiang, Chenyuan Wang, Cesar Andres Stuardo Moraga, Juntao Zhao, Bin Jia, Chengye Li, Zhongkai Zhao, Shixiong Zhao, Tiantian Fan, Hanshi Sun, Wenlei Bao, Shixun Wu, Zhekun Zhang, Yanbo Liang, Li-wen Chang, Jun Wang, Cheng Li, Li Han, Heng Zhang, Zhenbo Sun, Bo Liu, Xiaonan Nie, Ru Zhang, Hao Gong, Zuquan Song, Yucheng Nie, Jiawei Wu, Hongpeng Guo, Xinyi Di
\end{itemize}

Equally important, we thank everyone on the TorchTitan team and Edward Z. Yang for the insightful discussions and collaboration within the open-source community.

\clearpage
\bibliographystyle{plainnat}
\bibliography{ref}

@misc{li2025vescale,
      title={{veScale: Consistent and Efficient Tensor Programming with Eager-Mode SPMD}}, 
      author={Youjie Li and Cheng Wan and Zhiqi Lin and Hongyu Zhu and Jiacheng Yang and Ziang Song and Xinyi Di and Jiawei Wu and Huiyao Shu and Wenlei Bao and Yanghua Peng and Haibin Lin and Li-Wen Chang},
      year={2025},
      eprint={2509.07003},
      archivePrefix={arXiv},
      primaryClass={cs.PL},
      url={https://arxiv.org/abs/2509.07003}, 
}

@misc{ma2025veomni,
      title={{VeOmni: Scaling Any Modality Model Training with Model-Centric Distributed Recipe Zoo}}, 
      author={Qianli Ma and Yaowei Zheng and Zhelun Shi and Zhongkai Zhao and Bin Jia and Ziyue Huang and Zhiqi Lin and Youjie Li and Jiacheng Yang and Yanghua Peng and Zhi Zhang and Xin Liu},
      year={2025},
      eprint={2508.02317},
      archivePrefix={arXiv},
      primaryClass={cs.CL},
      url={https://arxiv.org/abs/2508.02317}, 
}

@article{kaplan2020scaling,
  title={Scaling laws for neural language models},
  author={Kaplan, Jared and McCandlish, Sam and Henighan, Tom and Brown, Tom B and Chess, Benjamin and Child, Rewon and Gray, Scott and Radford, Alec and Wu, Jeffrey and Amodei, Dario},
  journal={arXiv preprint arXiv:2001.08361},
  year={2020}
}

@article{wen2025fantastic,
  title={Fantastic Pretraining Optimizers and Where to Find Them},
  author={Wen, Kaiyue and Hall, David and Ma, Tengyu and Liang, Percy},
  journal={arXiv preprint arXiv:2509.02046},
  year={2025}
}

@article{wu2025terabyte,
  title={Terabyte-Scale Analytics in the Blink of an Eye},
  author={Wu, Bowen and Cui, Wei and Curino, Carlo and Interlandi, Matteo and Sen, Rathijit},
  journal={arXiv preprint arXiv:2506.09226},
  year={2025}
}

@article{zhao2023pytorch,
  title={Pytorch fsdp: experiences on scaling fully sharded data parallel},
  author={Zhao, Yanli and Gu, Andrew and Varma, Rohan and Luo, Liang and Huang, Chien-Chin and Xu, Min and Wright, Less and Shojanazeri, Hamid and Ott, Myle and Shleifer, Sam and others},
  journal={arXiv preprint arXiv:2304.11277},
  year={2023}
}

@article{agarwal2025gpt,
  title={gpt-oss-120b \& gpt-oss-20b model card},
  author={Agarwal, Sandhini and Ahmad, Lama and Ai, Jason and Altman, Sam and Applebaum, Andy and Arbus, Edwin and Arora, Rahul K and Bai, Yu and Baker, Bowen and Bao, Haiming and others},
  journal={arXiv preprint arXiv:2508.10925},
  year={2025}
}

@article{xu2021gspmd,
  title={Gspmd: general and scalable parallelization for ml computation graphs},
  author={Xu, Yuanzhong and Lee, HyoukJoong and Chen, Dehao and Hechtman, Blake and Huang, Yanping and Joshi, Rahul and Krikun, Maxim and Lepikhin, Dmitry and Ly, Andy and Maggioni, Marcello and others},
  journal={arXiv preprint arXiv:2105.04663},
  year={2021}
}

@article{smith2022using,
  title={Using deepspeed and megatron to train megatron-turing nlg 530b, a large-scale generative language model},
  author={Smith, Shaden and Patwary, Mostofa and Norick, Brandon and LeGresley, Patrick and Rajbhandari, Samyam and Casper, Jared and Liu, Zhun and Prabhumoye, Shrimai and Zerveas, George and Korthikanti, Vijay and others},
  journal={arXiv preprint arXiv:2201.11990},
  year={2022}
}

@article{dubey2024llama,
  title={The llama 3 herd of models},
  author={Dubey, Abhimanyu and Jauhri, Abhinav and Pandey, Abhinav and Kadian, Abhishek and Al-Dahle, Ahmad and Letman, Aiesha and Mathur, Akhil and Schelten, Alan and Yang, Amy and Fan, Angela and others},
  journal={arXiv e-prints},
  pages={arXiv--2407},
  year={2024}
}

@article{shoeybi2019megatron,
  title={Megatron-lm: Training multi-billion parameter language models using model parallelism},
  author={Shoeybi, Mohammad and Patwary, Mostofa and Puri, Raul and LeGresley, Patrick and Casper, Jared and Catanzaro, Bryan},
  journal={arXiv preprint arXiv:1909.08053},
  year={2019}
}

@inproceedings{rajbhandari2020zero,
  title={Zero: Memory optimizations toward training trillion parameter models},
  author={Rajbhandari, Samyam and Rasley, Jeff and Ruwase, Olatunji and He, Yuxiong},
  booktitle={SC20: International Conference for High Performance Computing, Networking, Storage and Analysis},
  pages={1--16},
  year={2020},
  organization={IEEE}
}

@inproceedings{jiang2024megascale,
  title={$\{$MegaScale$\}$: Scaling large language model training to more than 10,000 $\{$GPUs$\}$},
  author={Jiang, Ziheng and Lin, Haibin and Zhong, Yinmin and Huang, Qi and Chen, Yangrui and Zhang, Zhi and Peng, Yanghua and Li, Xiang and Xie, Cong and Nong, Shibiao and others},
  booktitle={21st USENIX Symposium on Networked Systems Design and Implementation (NSDI 24)},
  pages={745--760},
  year={2024}
}

@article{team2024gemini,
  title={Gemini 1.5: Unlocking multimodal understanding across millions of tokens of context},
  author={Team, Gemini and Georgiev, Petko and Lei, Ving Ian and Burnell, Ryan and Bai, Libin and Gulati, Anmol and Tanzer, Garrett and Vincent, Damien and Pan, Zhufeng and Wang, Shibo and others},
  journal={arXiv preprint arXiv:2403.05530},
  year={2024}
}

@inproceedings{gupta2018shampoo,
  title={Shampoo: Preconditioned stochastic tensor optimization},
  author={Gupta, Vineet and Koren, Tomer and Singer, Yoram},
  booktitle={International Conference on Machine Learning},
  pages={1842--1850},
  year={2018},
  organization={PMLR}
}

@article{team2025kimi,
  title={Kimi k2: Open agentic intelligence},
  author={Team, Kimi and Bai, Yifan and Bao, Yiping and Chen, Guanduo and Chen, Jiahao and Chen, Ningxin and Chen, Ruijue and Chen, Yanru and Chen, Yuankun and Chen, Yutian and others},
  journal={arXiv preprint arXiv:2507.20534},
  year={2025}
}

@TechReport{pytorch2024fsdp2,
  author = 	 "PyTorch",
  title = 	 "Fully Sharded Data Parallel (FSDP2)",
  howpublished = {\url{https://docs.pytorch.org/tutorials/intermediate/FSDP_tutorial.html}},
  year = 	 "2024",
}

@misc{jordan2024muon,
  author       = {Keller Jordan and Yuchen Jin and Vlado Boza and Jiacheng You and
                  Franz Cesista and Laker Newhouse and Jeremy Bernstein},
  title        = {Muon: An optimizer for hidden layers in neural networks},
  year         = {2024},
  url          = {https://kellerjordan.github.io/posts/muon/}
}

@misc{per_parameter_shard_rfc,
    author = {Gu, Andrew and Feng, Wei and Zhao, Yanli},
    title = {[RFC] Per-Parameter-Sharding FSDP},
    url = {https://github.com/pytorch/pytorch/issues/114299},
    year = 	 "2023",
}

@article{garey1975complexity,
  title={Complexity results for multiprocessor scheduling under resource constraints},
  author={Garey, Michael R and Johnson, David S.},
  journal={SIAM journal on Computing},
  volume={4},
  number={4},
  pages={397--411},
  year={1975},
  publisher={SIAM}
}

@misc{deepspeed_AG,
    author = {Halilakin},
    title = {DeepSpeed is slower than FSDP},
    url = {https://github.com/deepspeedai/DeepSpeed/issues/5047#issuecomment-1926275502},
    year={2024},
}

@misc{fsdp_record_stream,
  author = {Xu, Jane},
  title  = {{FSDP} \& {CUDACachingAllocator}},
  howpublished = {\url{https://dev-discuss.pytorch.org/t/fsdp-cudacachingallocator-an-outsider-newb-perspective/1486}},
  note   = {PyTorch Dev Discuss},
  year   = {2024},
}

@misc{fsdp1_reduce,
    author = {Zhao, Yanli and Gu, Andrew and Varma, Rohan and Luo, Liang and Huang, Chien-Chin and Xu, Min and Wright, Less and Shojanazeri, Hamid and Ott, Myle and Shleifer, Sam and others},
    title = {FSDP1 post backward reduce},
    url = {https://github.com/pytorch/pytorch/blob/a4925c0ce004cf883fdd1b248d71676769524934/torch/distributed/fsdp/_runtime_utils.py#L695C1-L773C1},
    year={2025},
}

@TechReport{megatron_fsdp,
  author = 	 "Megatron",
  title = 	 "MCore Custom Fully Sharded Data Parallel (FSDP)",
  howpublished = {\url{https://docs.nvidia.com/megatron-core/developer-guide/0.15.0/api-guide/custom_fsdp.html}},
  year = 	 "2025",
}

@article{liu2024deepseek,
  title={Deepseek-v3 technical report},
  author={Liu, Aixin and Feng, Bei and Xue, Bing and Wang, Bingxuan and Wu, Bochao and Lu, Chengda and Zhao, Chenggang and Deng, Chengqi and Zhang, Chenyu and Ruan, Chong and others},
  journal={arXiv preprint arXiv:2412.19437},
  year={2024}
}

@article{li2020pytorch,
  title={Pytorch distributed: Experiences on accelerating data parallel training},
  author={Li, Shen and Zhao, Yanli and Varma, Rohan and Salpekar, Omkar and Noordhuis, Pieter and Li, Teng and Paszke, Adam and Smith, Jeff and Vaughan, Brian and Damania, Pritam and others},
  journal={arXiv preprint arXiv:2006.15704},
  year={2020}
}

@misc{jaggedtensor,
    author = {PyTorch},
    title = {PyTorch JaggedTensor},
    url = {https://docs.pytorch.org/FBGEMM/fbgemm_gpu/overview/jagged-tensor-ops/JaggedTensorOps.html},
    year = 	 "2025",
}

@misc{nestedtensor,
    author = {PyTorch},
    title = {PyTorch NestedTensor},
    url = {https://docs.pytorch.org/docs/main/nested.html},
    year = 	 "2025",
}

@misc{raggedtensors,
    author = {TensorFlow},
    title = {TensorFlow Ragged Tensors},
    url = {https://www.tensorflow.org/guide/ragged_tensor},
    year = 	 "2025",
}

@inproceedings{dettmers8,
  title={8-bit Optimizers via Block-wise Quantization},
  author={Dettmers, Tim and Lewis, Mike and Shleifer, Sam and Zettlemoyer, Luke},
  booktitle={International Conference on Learning Representations},
  year={2022}
}

@misc{nccl16byte,
    author = {NVIDIA NCCL},
    title = {Regarding the AllGather bandwidth with different byte alignment under different protocols},
    url = {https://github.com/NVIDIA/nccl/issues/413},
    year = 	 "2025",
}

@misc{nccl2025collective,
    author = {NVIDIA NCCL},
    title = {NCCL: Collective Operations},
    url = {https://docs.nvidia.com/deeplearning/nccl/user-guide/docs/usage/collectives.html},
    year = 	 "2025",
}

@article{lepikhin2020gshard,
  title={Gshard: Scaling giant models with conditional computation and automatic sharding},
  author={Lepikhin, Dmitry and Lee, HyoukJoong and Xu, Yuanzhong and Chen, Dehao and Firat, Orhan and Huang, Yanping and Krikun, Maxim and Shazeer, Noam and Chen, Zhifeng},
  journal={arXiv preprint arXiv:2006.16668},
  year={2020}
}

@misc{dcp,
 author = {{PyTorch Team}},
 title = {Distributed Checkpoint},
 url = {https://docs.pytorch.org/docs/stable/distributed.checkpoint.html#distributed-checkpoint-torch#-distributed-checkpoint},
year = "2025"
}

@misc{pytorch_2026_h1_roadmap,
 author = {{PyTorch Team}},
 title = {Meta PyTorch Team 2026 H1 Roadmaps},
 url = {https://dev-discuss.pytorch.org/t/meta-pytorch-team-2026-h1-roadmaps},
 year = "2026"
}

@Misc{torch-dtensor,
  author = {{The PyTorch Team}},
  year = {2024},
  title = {{PyTorch DTensor (Distributed Tensor)}},
  howpublished = {\url{https://pytorch.org/docs/stable/distributed.tensor.html}},
}

\end{document}